\documentclass[11pt,english]{article}
\usepackage{jheppub}
\usepackage{dsfont}
\usepackage{graphicx}
\usepackage{amssymb}
\usepackage{subcaption}
\usepackage{amsmath}
\usepackage{slashed}
\usepackage{hyperref}
\usepackage{caption}
\usepackage[utf8]{inputenc}
\usepackage{xcolor}
\usepackage{babel}
\usepackage{array}

\newcommand\mc[1]{{\mathcal{#1}}}

\newcommand\CD{{\mc{D}}}
\newcommand\CL{{\mc{L}}}

\newcommand{\Tr}{{\rm Tr\,}}

\newcommand\Dslash{D\!\!\!\!\slash}

\def\shrug{\texttt{\raisebox{0.75em}{\char`\_}\char`\\\char`\_\kern-0.5ex(\kern-0.25ex\raisebox{0.25ex}{\rotatebox{45}{\raisebox{-.75ex}"\kern-1.5ex\rotatebox{-90})}}\kern-0.5ex)\kern-0.5ex\char`\_/\raisebox{0.75em}{\char`\_}}}

\title{Master 3d Bosonization Duality with Boundaries}

\preprint{\today}

\author[a]{Kyle Aitken,}
\author[a]{Andreas Karch,}
\author[b]{ Brandon Robinson}
\affiliation[a]{Department of Physics, University of Washington, Seattle, WA 98195, USA}
\affiliation[b]{School of Physics \& Astronomy and STAG Research Centre, University of Southampton,
Highfield, Southampton SO17 1BJ, U. K.}

\abstract{We establish the action of the three-dimensional non-Abelian bosonization dualities in the presence of a boundary, which supports a non-anomalous two-dimensional theory. In particular, we generalize a prescriptive method for assigning duality consistent boundary conditions used originally for Abelian dualities to dual non-Abelian Chern-Simons-matter theories with $SU$ and $U$ gauge groups and fundamental matter sectors.  The cases of single species matter sectors and those with both scalars and fermions in the dual theories are considered. Generalization of our methods to $SO$ and $USp$ Chern-Simons theories is also discussed.}

\begin{document}
\maketitle

\section{Introduction}

In the past few years, there has been a resurgence in the interest paid to quantum field theories in 2+1 dimensions and the unexpected IR dualities that they can exhibit.  In particular, it has been shown in a variety of contexts that Chern-Simons theories coupled to matter participate in bosonization-type dualities \cite{Aharony:2015mjs, Hsin:2016blu, Aharony:2016jvv, Jensen:2017dso}, and in the Abelian Chern-Simons matter theories, bosonization can then be extended to a web of particle-vortex dualities \cite{Karch:2016sxi,Seiberg:2016gmd}.    Apart from these curious dualities, Chern-Simons matter theories at IR fixed points are models of critical phenomena in low dimensional condensed matter systems, capture phenomena such as the fractional quantum Hall effect, and play a role in topological quantum computing \cite{Nayak:2008zza}.  Further, there have been lessons taken from these dualities and imported back into theories closer to ones recognizable to high energy physicists \cite{Komargodski:2017keh, Gaiotto:2017tne}

More than being purely speculative relations, 2+1 dimensional bosonization and particle-vortex dualities are now backed by significant evidentiary support.  In the case of non-Abelian Chern-Simons theories coupled to fundamental matter, the most analytically controllable evidence comes from the large $N$, large Chern-Simons level, $k$, fixed number of flavors regime, e.g. \cite{Giombi:2017txg}.  Despite $N=1$ lying well outside of any extrapolation from the non-Abelian evidence, recent exact lattice \cite{Chen:2017lkr} and quantum wire constructions \cite{Mross:2015idy, Mross:2017gny} have given strong support of Abelian 2+1 bosonization and particle-vortex duality.  All of this evidence suggests that these dualities might play a role in real world physical systems at criticality.

However if there is any intent on exporting the lessons from 2+1 dimensional bosonization to inform any aspect of experimental protocol for real world samples, then it is important to understand how the dualities can be made consistent with the introduction of a boundary.    In \cite{Aitken:2017nfd}, it was shown that a prescriptive method exists to assign boundary conditions for Abelian bosonization and particle-vortex dualities.  The key realization in arriving at the correct accounting for necessary boundary conditions and edge modes in Abelian dualities was that the Chern-Simons terms were best thought of  in terms motivated by their UV origins.  That is, the Chern-Simons terms are replaced by a theory of heavy, well-regulated ``fiducial fermions''.

Recently, there have been proposals to explain the vast set of 2+1 dimensional bosonization dualities as originating from a single ``master'' duality \cite{Jensen:2017bjo, Benini:2017aed}.  The master dualities generalize the original non-Abelian bosonization dualities (see \cite{Aharony:2015mjs}) to a level-rank type equivalence between Chern-Simons matter theories with both scalars and fermions in each theory. In addition, studying the $N=k=1$ limit with single species of fundamental matter can be shown to recover the known Abelian dualities.

 At first glance, one might assume that the application of the fiducial fermion analysis in Abelian dualities in the presence of a boundary would be a trivial generalization to the non-Abelian dualities. However because both sides of the single species non-Abelian and master bosonization dualities in \cite{Aharony:2015mjs, Jensen:2017bjo, Benini:2017aed} have dynamical gauge fields, the introduction of a boundary is more subtle.  As pointed out in \cite{gaiotto}, choosing certain boundary conditions on bulk dynamical gauge fields can alter the counting of global symmetries present on the boundary.  The interplay between boundary conditions and global symmetries presents an interesting, non-trivial extension of the study in \cite{Aitken:2017nfd} and will require a careful analysis in order to find duality-consistent boundary conditions.

To the end of finding duality-consistent boundary conditions for the master bosonization duality, we will organize our work as follows.  In section~\ref{sec:prelims}, we will briefly review the master bosonization duality and the fiducial fermion prescription.  After reviewing the preliminary information, we will construct in section~\ref{sec:single-species-boundaries} the consistent boundary conditions and necessary edge modes for non-Abelian dualities with a single species of matter.  Using the method developed to address the single species non-Abelian dualities, we will show how the new prescription works in all massive phases of the master bosonization duality in section~\ref{sec:master-boundaries}. Finally, we briefly comment on generalization to the $SO$ and $USp$ versions of the master duality in section~\ref{sec:so and usp}.

\section{Preliminaries}\label{sec:prelims}
\subsection{Master Bosonization Duality}
At the core of recently discovered non-supersymmetric 2+1 dimensional dualities is the well-known level-rank duality,
\begin{equation}
SU(N)_{-k}\leftrightarrow U(k)_{N}+U\left(kN\right)_{-1}\label{eq:level-rank}
\end{equation}
where $U\left(kN\right)_{-1}$ represents a gravitational Chern-Simons term $+2kN\text{CS}_\text{grav}$ \cite{Hsin:2016blu}.  New work has suggested that there exists a master duality that seemingly encompasses and generalizes all known 2+1 dimensional dualities built on the level-rank core \cite{Benini:2017aed,Jensen:2017bjo}. Schematically, the master duality is given by an equivalence between two gauge theories with fundamental matter sectors containing $N_s$ ($N_f$) scalars, $\phi$ ($\Phi$), and $N_f$ ($N_s$) Dirac fermions, $\psi$ ($\Psi$), i.e.
\begin{equation}
SU(N)_{-k+\frac{N_{f}}{2}}\,\text{with \ensuremath{N_{s}} \ensuremath{\phi}\ and \ensuremath{N_{f}} \ensuremath{\psi}}\qquad\leftrightarrow\qquad U(k)_{N-\frac{N_{s}}{2}}\,\text{with \ensuremath{N_{f}} \ensuremath{\Phi}\ and \ensuremath{N_{s}} \ensuremath{\Psi}}.\label{eq:mdb master schematic}
\end{equation}
This is the ``master'' duality in the sense that it together with its time-reversed version encompass all of the $3d$ bosonization dualities of ref. \cite{Aharony:2015mjs}. Namely, the single species limits of either $N_{f}=0$ or $N_{s}=0$ respectively yield
\begin{align}
SU(N)_{-k}\,\text{with \ensuremath{N_{s}} \ensuremath{\phi}}\qquad & \leftrightarrow\qquad U(k)_{N-\frac{N_{s}}{2}}\,\text{with \ensuremath{N_{s}} \ensuremath{\Psi}},\\
SU(N)_{-k+\frac{N_{f}}{2}}\,\text{with \ensuremath{N_{f}} \ensuremath{\psi}}\qquad & \leftrightarrow\qquad U(k)_{N}\,\text{with \ensuremath{N_{f}} \ensuremath{\Phi}}.
\end{align}

Explicitly, while starting from different UV theories, (\ref{eq:mdb master schematic}) is a duality between the partition functions
\begin{equation}
\int \mathcal{D}(\cdots)\;e^{-\int d^3x\;\CL_{SU}} \qquad \leftrightarrow \qquad \int \mathcal{D}(\cdots)\;e^{-\int d^3x\;\CL_{U}}
\end{equation}
in the IR limit.  The Lagrangians are given by\footnote{These Lagrangians are based on those in \cite{Jensen:2017bjo} with $\tilde{A}_{1}\to N\tilde{A}_{1}$ for simplicity of the expressions. This amounts to saying quarks have charge 1 under $U(1)_m$ rather than baryons.}
\begin{align}
\hspace{-0.2cm}\CL_{SU} & =\left|D_{b+B}\phi\right|^{2}+i\bar{\psi}\Dslash_{b+C-\tilde{A}_{2}}\psi-i\left((N_f-k)\text{CS}_N[b]+\text{BF}\left[f;\text{Tr}_{N}\left(b-\mathds{1}_N\left(\tilde{A}_{1}+\tilde{A}_{2}\right)\right)\right]\right)\nonumber \\
\hspace{-0.2cm} &\qquad -iN\left(\text{CS}_{N_{f}}[C]+\left(k-N_{f}\right)\left( \text{BF}[\tilde{A}_{1};\tilde{A}_{2}]+\text{CS}_1[\tilde{A}_{2}]\right)+2N_{f}\text{CS}_\text{grav}\right)+\CL_{\text{int}},\label{eq:mdb lsu}\\
\hspace{-0.2cm}\CL_{U} & =\left|D_{c+C}\Phi\right|^{2}+i\bar{\Psi}\Dslash_{c+B-\tilde{A}_{2}}\Psi-iN\left(\text{CS}_{k}[c]+\text{BF}[\text{Tr}_{k}\left(c\right);\tilde{A}_{1}]+2k\text{CS}_\text{grav}\right)+\CL_{\text{int}}^{\prime},\label{eq:mdb lu}
\end{align}
where the $U(1)$ field $f$ is a Lagrange multiplier whose precise role -- as well as the motivation for the notation adopted for the gauge fields listed in Table~\ref{tab:notation} -- will be discussed below. For brevity, we have adopted the following notation for Chern-Simons and BF terms for rank $N$ gauge groups
\begin{align}
\text{CS}_{N}[b] & \equiv\frac{1}{4\pi}\text{Tr}_{N}\left(bdb-i\frac{2}{3}b^{3}\right),\\
\text{BF}[f;\text{Tr}_{N}b] & \equiv\frac{1}{2\pi}fd\text{Tr}_{N}b.
\end{align}
The gravitational Chern-Simons term is given by
\begin{equation}
\int_{\mathcal{M}=\partial X}\text{CS}_\text{grav}\equiv\frac{1}{192\pi}\int_{X}\text{Tr}R\wedge R,
\end{equation}
where $X$ is a $d=4$ manifold and $\mathcal{M}$ is its $d=3$ boundary.

The interactions terms represent all possible relevant and marginal operators consistent with symmetries\footnote{Although the mixed scalar and fermion interactions are marginal in the IR at leading order in the large $N$ limit, the sign of the subleading corrections are currently unknown. As in \cite{Jensen:2017bjo,Benini:2017aed} we will assume such operators are at least marginal since they are vital for the consistency of the master duality. }
\begin{align}
\CL_{\text{int}} & =\alpha_\varphi(\phi^{\dagger\alpha M}\phi_{\alpha M})^{2}-(\bar{\psi}^{\alpha I}\phi_{\alpha M})(\phi^{\dagger\beta M}\psi_{\beta I})\label{eq:mdb int 1}\\
\CL_{\text{int}}^{\prime} & =\alpha_\varphi(\Phi^{\dagger\rho I}\Phi_{\rho I})^{2}+(\bar{\Psi}^{\rho M}\Phi_{\rho I})(\Phi^{\dagger\sigma I}\Psi_{\sigma M})\label{eq:mdb int 2}
\end{align}
where $\alpha,\beta=1,\ldots,N$ and $\rho,\sigma=1,\ldots,k$ are color labels, $I=1,\ldots,N_{f}$ and $M=1,\ldots,N_{s}$ are flavor labels, and $\alpha_\varphi$ is the scalar self-coupling that is tuned to $\alpha_\varphi\rightarrow\infty$ at the IR fixed point. In what follows we will often drop the explicit indices and denote the interaction terms by, e.g., $|\phi|^4$ and $\bar{\Psi}\Psi|\Phi|^2$.

The $|\phi|^4$ and $|\Phi|^4$ are the usual interactions which are present at the Wilson-Fischer fixed point. The effect of the scalar-fermion mixing term is to give a subset of the $N_f$ (or $N_s$) fermions a mass when the scalars in the theory acquire a nonzero vacuum expectation value. The additional effect of this mass from the mixing term is necessary to get complete agreement between the two sides of the duality, and the relative sign between the mixing terms in $\CL_{\text{int}}$ and $\CL_{\text{int}}^\prime$ is important to match the phases.

\begin{table}
\begin{centering}
\begin{tabular}{|c|c|c|c|c||c|c|c|c|}
\cline{2-9}
\multicolumn{1}{c|}{}& \multicolumn{4}{c||}{Gauge Fields} & \multicolumn{4}{c |}{Background Fields} \\
\hline
\textbf{Symmetry} &$U(N)$& $U(k)$ & $SU(N)$ & $SU(k)$ &$SU(N_s)$ & $SU(N_f)$& $U(1)_{m,b}$& $U(1)_{F,S}$
\tabularnewline
\hline
\textbf{Field} & $b_\mu$ & $c_\mu$ & $b'_{\mu}$& $c'_{\mu}$ & $B_\mu$ & $C_\mu$& $\tilde{A}_{1\mu}$&  $\tilde{A}_{2\mu}$
\tabularnewline
\hline
\textbf{Index} & $\alpha$, $\beta$ & $\rho$, $\sigma$ & $\alpha$, $\beta$ & $\rho$, $\sigma$ & $M$, $N$ &
$I$, $J$ &{\bf{---}{---}} & {\bf{---}{---}} \\
\hline
\end{tabular}
\par\end{centering}
\caption{Collection of notation for various gauge fields.  Note that dynamical gauge fields are indicated by lower case letters and background gauge fields by upper case letters. \label{tab:notation}}
\end{table}

In what follows, we will denote dynamical gauge fields by lowercase letters and background gauge fields by uppercase. Ordinary gauge connections will be denoted by $b, B, c, C$ and spin$_c$ connections by $A, a$.\footnote{For a review on the subtleties of spin$_c$ connections in the context of these dualities, see \cite{Seiberg:2016gmd}.} Specifically, we have denoted by $b_{\mu}$ a dynamical $U(N)$ gauge field, $c_{\mu}$ a dynamical $U(k)$ gauge field, $C_{\mu}$ a background $SU(N_{f})$ gauge field, and $B_{\mu}$ a background $SU(N_{s})$ gauge field.  Further, the background spin$_c$ gauge fields for  $U(1)_{m,b}$ and $U(1)_{F,S}$ are respectively $\tilde{A}_{1\mu}$ and $\tilde{A}_{2\mu}$, and $f_{\mu}$ is a dynamical $U(1)$ field, which acts as a Lagrange multiplier. The covariant derivatives are given by
\begin{subequations}
\begin{align}
(D_{b+B})_\mu\phi & =\left[\partial_{\mu}-i\left(b_{\mu}\mathds{1}_{N_{s}}+B_{\mu}\mathds{1}_{N}\right)\right]\phi,\\
(D_{b+C-\tilde{A}_{2}})_\mu\psi & =\left[\partial_{\mu}-i\left(b_{\mu}\mathds{1}_{N_{f}}+C_{\mu}\mathds{1}_{N}-\tilde{A}_{2\mu}\mathds{1}_{NN_{f}}\right)\right]\psi,\\
(D_{c+C})_\mu\Phi & =\left[\partial_{\mu}-i\left(c_{\mu}\mathds{1}_{N_{f}}+C_{\mu}\mathds{1}_{k}\right)\right]\Phi,\\
(D_{c+B-\tilde{A}_{2}})_\mu\Psi & =\left[\partial_{\mu}-i\left(c_{\mu}\mathds{1}_{N_{s}}+B_{\mu}\mathds{1}_{k}-\tilde{A}_{2\mu}\mathds{1}_{kN_{s}}\right)\right]\Psi.
\end{align}
\end{subequations}
$\mathds{1}_n$ is the $n$-dimensional identity matrix. Although the presence of the Lagrange multiplier $f$ makes coupling slightly obscure, on the $U$ side $\tilde{A}_{1}$ only appears through a BF coupling to the monopole current $\star j_{m}=\frac{1}{2\pi}d\text{Tr}_{k}c$, while on the $SU$ side it couples directly to the particle number current. The $\tilde{A}_2$ field is associated with a new symmetry which arises due to the presence of both scalars and fermions on each side of the duality. With the Lagrange multiplier, the $U(1)_{F,S}$ symmetry only couples to the fermions on each side of the duality, although once $f$ is integrated out it couples only to $\phi$ on the $SU$ side.

In the way that we have written the fermions in (\ref{eq:mdb lsu}) and (\ref{eq:mdb lu}), we have left implicit the regularizing $\eta$-invariant terms for the Dirac fermions \cite{Witten:2015aba,Witten:2015aoa}. This is the notation established in \cite{Seiberg:2016gmd}.  Being very explicit, for $N_{f}$, $N$-component fermions we have absorbed into the kinetic term what is often written as a half-integer Chern-Simons term that results from integrating out heavy regulator fermions, i.e.
\begin{equation}
i\bar{\psi}\Dslash_{b}\psi-i\left[-\frac{N_{f}}{8\pi}\text{Tr}_{N}\left(bdb-i\frac{2}{3}b^{3}\right)\right]\qquad\xrightarrow{\hspace{1cm}}\qquad i\bar{\psi}\Dslash_{b}\psi.
\end{equation}
This convention is chosen such that when integrating out positive mass dynamical fermions the hidden $\eta$-invariant term is canceled, which leaves the Chern-Simons levels unchanged. However, when a negative mass fermion is integrated out the overall effect is to shift the associated Chern-Simons levels by $N_f$. This will be the convention we use for fermions throughout this paper.\footnote{This will slightly complicate things when we time-reverse the duality, because this transformation should also flip the $\eta$-invariant term. However, we will keep the same convention whether or not we are talking about the original or time-reversed duality. The net effect of this will mean time-reversal comes with a shift in Chern-Simons terms as well.}

As we mentioned above, the $SU$ side of the theory contains a Lagrange multiplier field $f$, which effectively transforms $SU(N)\to U(N)\times U(1)$. Analyzing the symmetry breaking pattern for $U(N)\times U(1)$ is easier than for $SU(N)$ \cite{Benini:2017aed, Jensen:2017bjo, Hsin:2016blu}. Occasionally, it will be useful to look at the original $SU$ Lagrangian with $f$ integrated out,
\begin{align}\label{eq:su_prime}
\CL_{SU}^{\prime} & =\left|D_{b^{\prime}+B+\tilde{A}_{1}+\tilde{A}_{2}}\phi\right|^{2}+i\bar{\psi}\Dslash_{b^{\prime}+C+\tilde{A}_{1}}\psi-i\left((N_f-k)\text{CS}_{N}[b^{\prime}]+N\text{CS}_{N_f}[C]\right)\nonumber \\
 & -i\left(-N(k-N_f)\text{CS}_1[\tilde{A}_1]+2NN_{f}\text{CS}_\text{grav}\right)+\CL_{\text{int}}.
\end{align}

Because the duality exactly at the IR fixed point is between what are in general strongly coupled theories, the best evidence for validity of $3d$ bosonization dualities comes from gapped phases where the identification can be directly verified. The dictionary for mass terms across the master duality is given by \cite{Jensen:2017bjo}\footnote{Note the opposite convention appears in \cite{Benini:2017aed} since it is the time-reversed version of the duality considered in \cite{Jensen:2017bjo}.}
\begin{equation}
m_{\psi}\leftrightarrow-m_{\Phi}^{2},\qquad m_{\phi}^{2}\leftrightarrow m_{\Psi}.
\end{equation}
Since we have two types of matter on each side of the duality, naively one would expect there to be four different mass-deformed phases. However, it has been shown that the interactions of (\ref{eq:mdb int 1}) and (\ref{eq:mdb int 2}) separate one of these four phases into two separate phases, giving us five phases total \cite{Benini:2017aed, Jensen:2017bjo}. Specifically, when the scalar acquires a vacuum expectation value the interactions give the so-called ``singlet fermions'', which are neutral under the unbroken gauge group, a mass shift.

\begin{figure}
\begin{centering}
\includegraphics[scale=0.55]{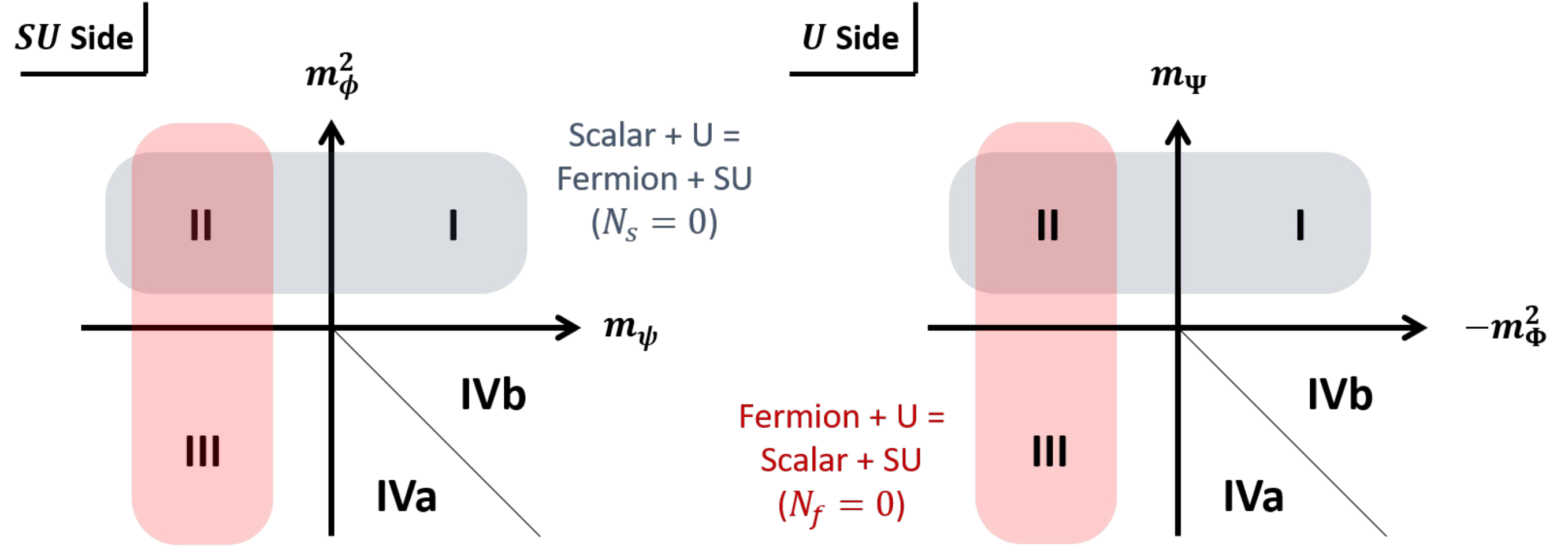}
\par\end{centering}
\caption{Various phases of the master duality and the non-Abelian reductions.
The shaded red and blue correspond to the single-matter non-Abelian
dualities.\label{fig:Various-phases-of master mdb}}
\end{figure}

The five massive phases are shown in Fig. \ref{fig:Various-phases-of master mdb}. On the $SU$ side we expect to find
\begin{subequations}
\begin{align}
\text{I}\,:\, & [SU(N)_{-k+N_{f}}\times U\left(NN_{f}\right)_{-1}]\times SU\left(N_{f}\right)_{N}\times SU\left(N_{s}\right)_{0}\times J_{I},\\
\text{II}\,:\, & [SU(N)_{-k}\times U\left(0\right)_{-1}]\times SU\left(N_{f}\right)_{0}\times SU\left(N_{s}\right)_{0}\times J_{II},\\
\text{III}\,:\, & [SU\left(N-N_{s}\right)_{-k}\times U\left(0\right)_{-1}]\times SU\left(N_{f}\right)_{0}\times SU\left(N_{s}\right)_{-k}\times J_{III},\\
\text{IVa}\,:\, & [SU\left(N-N_{s}\right)_{-k+N_{f}}\times U\left(N_{f}\left(N-N_{s}\right)\right)_{-1}]\nonumber\\
& \times SU\left(N_{f}\right)_{N-N_{s}}\times SU\left(N_{s}\right)_{-k}\times J_{IVa},\\
\text{IVb}\,:\, & [SU\left(N-N_{s}\right)_{-k+N_{f}}\times U\left(NN_{f}\right)_{-1}]\times SU\left(N_{f}\right)_{N}\times SU\left(N_{s}\right)_{-k+N_{f}}\times J_{IVb}.
\end{align}
\end{subequations}
Meanwhile, on the $U$ side,
\begin{subequations}
\begin{align}
\text{I}\,:\, & [U\left(k-N_{f}\right)_{N}\times U\left(kN\right)_{-1}]\times SU\left(N_{f}\right)_{N}\times SU\left(N_{s}\right)_{0}\times J_{I^\prime},\\
\text{II}\,:\, & [U(k)_{N}\times U\left(kN\right)_{-1}]\times SU\left(N_{f}\right)_{0}\times SU\left(N_{s}\right)_{0}\times J_{II^\prime},\\
\text{III}\,:\, & [U(k)_{N-N_{s}}\times U\left(k\left(N-N_{s}\right)\right)_{-1}]\times SU\left(N_{f}\right)_{0}\times SU\left(N_{s}\right)_{-k}\times J_{III^\prime},\\
\text{IVa}\,:\, & [U\left(k-N_{f}\right)_{N}\times U\left(k\left(N-N_{s}\right)\right)_{-1}]\times SU\left(N_{f}\right)_{N-N_{s}}\times SU\left(N_{s}\right)_{-k}\times J_{IVa^\prime},\\
\text{IVb}\,:\, & [U\left(k-N_{f}\right)_{N-N_{s}}\times U\left(kN+(N_{f}-k)N_{s}\right)_{-1}]\nonumber\\
& \times SU\left(N_{f}\right)_{N}\times SU\left(N_{s}\right)_{-k+N_{f}}\times J_{IVb^\prime}.
\end{align}
\end{subequations}
The bracketed are level-rank dual by \eqref{eq:level-rank}, while the rest of the terms are global symmetries and should be the same on both sides. The Abelian factors unique to each phase are given by
\begin{equation}
J_{i}\equiv J_{i}^{ab}\frac{1}{4\pi}\tilde{A}_{a}d\tilde{A}_{b}
\end{equation}
with $a,b=1,2$, $i$ indexing the phase $\{I,\ldots,IVb\}$, and
\begin{subequations}
\begin{align}
J_{I}^{ab} & =\left(\begin{array}{cc}
-N\left(k-N_{f}\right) & 0\\
0 & 0
\end{array}\right)\\
J_{II}^{ab} & =\left(\begin{array}{cc}
-Nk & 0\\
0 & 0
\end{array}\right)\\
J_{III}^{ab} & =-\frac{Nk}{N-N_{s}}\left(\begin{array}{cc}
N & N_{s}\\
N_{s} & N_{s}
\end{array}\right) \label{eq:phaseIII bkg}\\
J_{IVa}^{ab} & =-\frac{N\left(k-N_{f}\right)}{N-N_{s}}\left(\begin{array}{cc}
N & N_{s}\\
N_{s} & N_{s}
\end{array}\right)-N_{f}N_{s}\left(\begin{array}{cc}
0 & 0\\
0 & 1
\end{array}\right)\\
J_{IVb}^{ab} & =-\frac{N\left(k-N_{f}\right)}{N-N_{s}}\left(\begin{array}{cc}
N & N_{s}\\
N_{s} & N_{s}
\end{array}\right).
\end{align}
\end{subequations}
Since the massive phases are dual to one another, this is taken as good evidence that the master duality remains true at the conformal fixed point. A similar matching can be performed on the five critical lines that separate the five phases \cite{Jensen:2017bjo, Benini:2017aed}.

\subsection{Adding Boundaries}

For simplicity, we will consider the theory on the half-space $\mathbb{\mathbb{R}}_{+}^{2,1}$ with coordinates $\left\{ t,x,y\right\} $ with $t,x\in\left(-\infty,\infty\right)$ and $y\geq0$. As in the Abelian duality, the results should be largely independent of the choice of $\mathbb{R}_+^{2,1}$ as our background \cite{Aitken:2017nfd}. We will use $i,j=\left\{x,t\right\}$ to refer to indices parallel to the boundary.

In this section we will start by briefly summarizing our conventions for boundary conditions for single-component fields as prescribed for Abelian dualities in \cite{Aitken:2017nfd}, which generalize fairly trivially to non-Abelian theories. Further, we will review the impact of the choice of boundary conditions on the presence of edge modes and anomalies in the boundary theory.  We will then review the method described in \cite{Aitken:2017nfd} for properly accounting for edge modes by introducing ``fiducial fermions''.

\subsubsection*{Boundary conditions}
From the perspective of the action, boundary conditions arise from partial integration and demanding a well-defined variational principle. The most basic conditions one encounters require either the variation of a dynamical variable  (``Dirichlet'') or its coefficient (``Neumann'') to vanish at the boundary.  Consistent boundary conditions for a scalar with non-derivative couplings can be either Neumann or Dirichlet,
\begin{equation}
(D_b)_{y}\phi_{\alpha M}\bigr|_{\partial}=0\qquad\text{or}\qquad\delta\phi_{\alpha M}\bigr|_{\partial}=0
\end{equation}
where ``$|_\partial$'' denotes an expression which holds at the boundary.
Equivalent boundary conditions hold for $\Phi_{\rho I}$. In order to derive the boundary conditions for a given Dirac fermion $\psi$, it is convenient to decompose $\psi$ into its left- and right-handed components, $\psi^\pm$,
\begin{equation}
\psi=\left(\begin{array}{c}
\psi^{+}\\
\psi^{-}
\end{array}\right),\qquad\text{i.e.}\quad\psi^{\pm}=P_{\pm}\psi.
\end{equation}
The projector $P_{\pm}=\left(\mathds{1}\pm\gamma^{y}\right)/2$ where $\gamma^{y}$ is the gamma matrix which is perpendicular to the boundary. The boundary conditions are then
\begin{equation}
\psi_{\alpha I}^{+}\bigr|_{\partial}=0\qquad\text{or}\qquad\psi_{\alpha I}^{-}\bigr|_{\partial}=0.
\end{equation}
Equivalent boundary conditions hold for $\Psi_{\rho M}$. The boundary conditions on the Pauli-Villars fields follow in an analogous manner. Like \cite{Aitken:2017nfd}, we chose boundary conditions for the Pauli-Villars that will never give rise to edge modes.

The boundary conditions for gauge fields also fall into the category of Neumann and Dirichlet boundary conditions,
\begin{align}
F_{iy}\bigr|_{\partial}=(\partial_y b_i - \partial_i b_y+[b_y,b_i])\bigr|_{\partial} =0,\qquad b_i\bigr|_{\partial}=0,
\end{align}
respectively.\footnote{Neumann boundary conditions can be modified by coupling boundary matter to the bulk gauge sector by $\epsilon^{ij} F_{jy}|_{\partial}=j^i_\text{bdry}$ where $j^i_\text{bdry}$ is the boundary matter current \cite{Dimofte:2017tpi}. Since we do not add any additional charged boundary matter, we will always set $j^i_\text{bdry}=0$.} In the Abelian dualities, there are only dynamical gauge fields on one side of the duality, and thus only one boundary condition is necessary \cite{Aitken:2017nfd}, which obviates the complications in choosing consistent boundary conditions in both theories. In this work, we will need to be more careful in choosing boundary conditions for all of the dynamical gauge fields.

For Neumann boundary conditions on dynamical gauge fields, we will need to worry about anomaly inflow. Note that since we do not assign boundary conditions for background fields, their corresponding Chern-Simons terms can produce anomalies. The cancellation of anomalies will be achieved by introducing ``fiducial fermions'', which will give rise to edge modes and will be discussed in the next section.

If we choose Dirichlet boundary conditions for the dynamical gauge fields, there is no chiral current flow off the boundary and, hence, no anomalies. This follows from the fact $j^y_\text{flux}|_{\partial}\sim F_{ij}|_{\partial}=0$.  Since Dirichlet boundary conditions break the gauge symmetry to the group that leaves the boundary condition invariant, an additional global symmetry emerges at the boundary \cite{Dimofte:2017tpi}.

We will show that the only way the global symmetries are consistent with the duality is to choose Dirichlet boundary conditions on one side and Neumann boundary conditions on the other. These results align with those discussed in \cite{gaiotto}.

Lastly, we should mention that choosing the same boundary condition on all flavors is necessary in order to maintain the full $SU\left(N_{s}\right)$ and $SU\left(N_{f}\right)$ global symmetries as well as the respective gauge symmetries. For future work, it may be interesting to consider a set of boundary conditions that breaks the flavor symmetries or mixing Neumann and Dirichlet boundary conditions for subsets of the gauge fields in a given theory.

\subsubsection*{Edge modes and anomalies}
In studying Chern-Simons matter theories in the presence of a boundary, we must reconcile the theories against possible edge modes allowed by the boundary conditions and any anomaly inflow.  In particular, introducing gapped fermions to a manifold with a boundary can create gapless, chiral fermionic modes localized to the boundary, i.e. domain wall fermions (DWFs).  If we allow the mass of the bulk fermions to vary in the direction normal to the boundary ($m(y)$), then by the standard construction \cite{Callan:1984sa,Kaplan:2009yg} DWFs will exist when the profile of the spatially varying mass leaves the function
\begin{align}
\xi^\pm(y) = e^{\pm\int_0^y dy^\prime m(y^\prime)}
\end{align}
finite for all $y\in \mathbb{R}^{2,1}_+$.  In fact in $\mathbb{R}^{2,1}_+$, {\emph{any}} constant, non-zero mass will give a normalizable zero mode with chirality determined by the sign of the mass.  That is, we have left-moving DWFs for sgn$(m)=+1$ and right-moving DWFs for sgn$(m)=-1$.

In addition to the possible anomalies associated with non-vanishing chiral currents on our boundary, we also need to take care of potential anomaly inflow from the gauge sector.  Chern-Simons theories in the presence of a boundary are not {\emph{a priori}} gauge invariant everywhere. However, the non-trivial anomaly associated with a bulk $SU(N)$ Chern-Simons term of level $k$ can be compensated for by the chiral anomaly through the Callan-Harvey mechanism provided \cite{Callan:1984sa}
\begin{align}
k = n_+-n_-,
\end{align}
where $k$ is the level of the bulk Chern-Simons theory and $n_\pm$ are the number of (right-) left-movers in the fundamental representation of $SU(N)$ living on the boundary. This of course generalizes to the Abelian case as well. Similarly the gravitational Chern-Simons term with coefficient $k_\Omega$ has an anomaly associated with diffeomorphisms, which can be compensated for by having excess right- or left-moving ($\tilde{n}_\pm$ resp.) Majorana-Weyl fermions satisfying
\begin{align}
k_{\Omega}= \frac{1}{2}(\tilde{n}_+ - \tilde{n}_-).
\end{align}
Equivalently, we could use a single right- or left-moving Weyl fermion for every two corresponding Majorana-Weyl fermions to accomplish the same compensation.

\subsubsection*{Fiducial fermions}
Informed by lattice realization of Abelian dualities, the accounting for edge modes above led to the prescriptive replacement of Chern-Simons terms by heavy fermions  \cite{Aitken:2017nfd}.  These ``fiducial fermions'' act to display the UV physics captured in the IR by the Chern-Simons terms while more directly enumerating the gauge sector edge modes.  The non-trivial IR theory left behind after integrating out heavy Dirac fermions coupled to a background spin$_c$ connection $A$ is CS$_{1}[A] + 2$CS$_\text{grav}$. More importantly, the fiducial fermions give rise to DWFs which automatically render their associated Chern-Simons terms non-anomalous.  Thus, the fiducial fermion prescription reads
\begin{equation}
e^{\pm i\int d^{3}x\,(\text{CS}_{1}[A]+2\text{CS}_\text{grav})}\rightarrow\int\CD\chi\CD\lambda\,e^{i\int d^{3}x\,\CL_{ff}^{\pm}[\chi,\lambda,A]},\label{eq:abelian ff}
\end{equation}
where
\begin{equation}
\CL_{ff}^{\pm}[\chi,\lambda,A]\equiv \lim_{|m_\chi|,|m_\lambda|\to\infty}\left(i\bar{\chi}\Dslash_{A}\chi\mp\left|m_{\chi}\right|\bar{\chi}\chi+i\bar{\lambda}\Dslash_{A}\lambda\mp\left|m_{\lambda}\right|\bar{\lambda}\lambda\right).
\end{equation}
Here $\chi$ is the fiducial fermion, $\lambda$ is the Pauli-Villars regulator field, and the their respective masses $|m_\chi|,\,|m_\lambda|$ are taken to be parametrically heavy.

This procedure generalizes to the case where $B$ is a non-Abelian background gauge field of $SU(N)$ and we have
\begin{equation}
e^{\pm i\int d^{3}x\,(k\text{CS}_{N}[B]+2Nk\text{CS}_\text{grav})}=\int\prod_{M=1}^{k}\CD\chi_{M}\CD\lambda_{M}\,e^{i\int d^{3}x\,\CL_{ff}^{\pm}[\chi_{M},\lambda_{M},B]},\label{eq:non-abelian ff}
\end{equation}
now with
\begin{equation}
\CL_{ff}^{\pm}[\chi_M,\lambda_M,B]\equiv\lim_{|m_{\chi_M}|,|m_{\lambda_M}|\to\infty}\left(i\bar{\chi}^M\Dslash_{B}\chi_{M}\mp\left|m_{\chi_{M}}\right|\bar{\chi}^{M}\chi_{M}+i\bar{\lambda}^{M}\Dslash_{B}\lambda_{M}\mp\left|m_{\lambda_{M}}\right|\bar{\lambda}^{M}\lambda_{M}\right)
\end{equation}
with $\chi_M$ and $\lambda_M$ in the fundamental representation of $SU(N)$. The non-Abelian fiducial fermion prescription requires $\chi_{M}$ and $\lambda_{M}$ be parametrically heavy $N$-component fields with $U(k)$ flavor symmetry.

As was done in \citep{Aitken:2017nfd}, it will be useful to rewrite all BF terms as Chern-Simons terms in order to properly account for the edge theories. For example,
\begin{align}
N\text{CS}_k [\tilde{c}] + N \text{BF}[\text{Tr}_k(\tilde{c});\tilde{A}_1\mathds{1}_{k}]=N\text{CS}_k [\tilde{c}+\tilde{A}_1\mathds{1}_{k}] - N \text{CS}_k[\tilde{A}_1\mathds{1}_{k}].
\end{align}
The right-hand side makes the assignments of fiducial fermions clearer.

\subsubsection*{Global symmetries}
The global symmetries manifest in the Lagrangian as three types of background Chern-Simons terms: (1) Abelian, namely $\tilde{A}_{1}$ and $\tilde{A}_{2}$, (2) non-Abelian, $B$ and $C$, and (3) gravitational. All three of these global symmetries are related to a conserved current, which will allow us to put additional constraints on the fields.

First consider the Abelian symmetries. There is an identification between the currents which couple to $\tilde{A}_a$, found via
\begin{align}
j_{U,a}^{\mu}(x)\equiv\frac{\delta S_{U}[\tilde{A}_{a}]}{\delta\tilde{A}_{a\mu}(x)},\qquad\leftrightarrow\qquad j_{SU,a}^{\mu}(x)\equiv\frac{\delta S_{SU}[\tilde{A}_{a}]}{\delta\tilde{A}_{a\mu}(x)}.
\end{align}
As with the Abelian dualities, $\tilde{A}_{1}$ is associated with the flux current on $U$ side and a particle current on the $SU$ side. For example, when $N_s=0$ and we set $\tilde{A}_1=0$ after variation,
\begin{align}
j_{U,1}^{\mu}=\frac{1}{2\pi}\epsilon^{\mu\nu\rho}\partial_{\nu}\text{Tr}_k(\tilde{c_\rho})\quad\leftrightarrow\quad j_{SU,1}^{\mu}=j_\text{fermion}^\mu.
\end{align}
We will show below that the $\tilde{A}_{2}$ field plays a very similar role. Note that in the single species non-Abelian dualities the $\tilde{A}_{2}$ symmetry drops out \cite{Aharony:2015mjs}, so it is only a feature of the master bosonization duality \cite{Jensen:2017bjo, Benini:2017aed}.

The non-Abelian global flavor symmetries also give two currents related to the $SU(N_{s})$ and $SU(N_{f})$ symmetries on either side.
These flavor currents are not just simply matter currents because there is also flux coupling to the background $C_{\mu}$ fields on the $SU$ side of the duality.

Lastly, the equivalence of the gravitational currents simply identifies the stress-energy tensors on either side of the duality. We will not make use of this identification in what follows.

\section{Single Species non-Abelian Bosonization with Boundaries}\label{sec:single-species-boundaries}

Before analyzing the master bosonization duality in the presence of a boundary, let us take the step of first considering non-Abelian dualities with a single species of matter. That is, we consider (\ref{eq:mdb lsu}) and (\ref{eq:mdb lu}) setting either $N_{f}=0$ or $N_{s}=0$, which correspond to one of Aharony's original dualities and one of the time-reversed versions \cite{Aharony:2015mjs}. Additionally, we make connections with the Abelian limit where we take $N_{s}=k=N=1$ or $N_{f}=k=N=1$ and find results consistent with our previous analysis in \cite{Aitken:2017nfd}.  We will also discuss additional subtleties involving the connections coupled to the fermion.

Setting either $N_s$ or $N_f=0$ eliminates one type of matter from each side of the master duality. This has the effect of making the additional $U(1)_{S,F}$ symmetry redundant.  Specifically, $U(1)_{S,F}$ becomes a linear combination of the global $U(1)_{m,b}$ symmetry and the dynamical gauge group. Since $U(1)_{S,F}$ does not appear in \cite{Aharony:2015mjs}, the redundancy should be expected. Importantly, $U(1)_{S,F}$ becoming redundant does not amount to just setting $\tilde{A}_2=0$ in the master duality. We will see keeping careful track of the $\tilde{A}_2$ dependence in Sec. \ref{sec:nab 2} allows us to correctly distinguish ordinary and spin$_c$ connections.

\subsection{Non-Abelian $U+\text{scalars}\leftrightarrow SU+\text{fermions}$}
\label{sec:nab 1}

To start studying the non-Abelian dualities, we will consider $N_{s}=0$. This reduces (\ref{eq:mdb master schematic}) to
\begin{align}
SU(N)_{-k+\frac{N_{f}}{2}}\,\text{with \ensuremath{N_{f}} \ensuremath{\psi}}\qquad & \leftrightarrow\qquad U(k)_{N}\,\text{with \ensuremath{N_{f}} \ensuremath{\Phi}}\label{eq:mdb ns0}
\end{align}
with the mass identification $m_{\psi}\leftrightarrow-m_{\Phi}^{2}$. This duality is subject to the flavor bound $N_{f}\leq k$.\footnote{There are arguments that these flavor bounds can be extended slightly \cite{Komargodski:2017keh}, but we will not consider such cases in this work.}

Explicitly, the Lagrangians for the theories on either side of (\ref{eq:mdb ns0}) are given by
\begin{align}
\CL_{SU}  =&i\bar{\psi}\Dslash_{\,b^\prime+C+\tilde{A}_1}\psi-i\left((N_{f}-k)\text{CS}_{N}[b^\prime]+N\text{CS}_{N_{f}}[C]\right) \nonumber \\
& -i\left(-N(k-N_f)\text{CS}_1[\tilde{A}_1]+2NN_{f}\text{CS}_\text{grav}\right)\label{eq:mdb s+f lsu}\\
\CL_{U}  =&\left|D_{c+C}\Phi\right|^{2}+\alpha_\varphi |\Phi|^4-i\left(N\text{CS}_{k}[c]+N\text{BF}[\text{Tr}_{k}\left(\tilde{c}\right);\tilde{A}_{1}]+2Nk\text{CS}_\text{grav}\right)\label{eq:mdb u+s lu}.
\end{align}
Since the Lagrange multiplier term will not be important for this section, we have integrated out $f$ as in \eqref{eq:su_prime}.  Furthermore, in \eqref{eq:mdb u+s lu}, we can split the $U(k)$ field, $c$, into its traceless $SU(k)$ part, $c^{\prime}$,
field and non-zero trace, $\tilde{c}$, such that
\begin{align}
\CL_U&=\left|D_{c+C}\Phi\right|^{2}+\alpha_\varphi |\Phi|^4-i\left(N\text{CS}_{k}[c^{\prime}]+N\text{CS}_{k}[\tilde{c}+\tilde{A}_{1}\mathds{1}_{k}]-Nk\text{CS}_{1}[\tilde{A}_{1}]+2Nk\text{CS}_\text{grav}\right)
\end{align}
Note that mass deformations in these theories correspond to phases I and II in Fig.~\ref{fig:Various-phases-of master mdb}. Specifically, $m_{\psi}<0$ and $m_{\Phi}^{2}>0$ is Phase II, and $m_{\psi}>0$ and $m_{\Phi}^{2}<0$ is Phase I. Also take note of the fact the duality has no $\tilde{A}_2$ dependence, since the $U(1)_{S,F}$ duality coupled to the fields associated with the $SU(N_s)$ symmetry.

Let us work through the counting of fiducial fermions in detail.  First consider the $U$ side of Phase II where $m_{\Phi}^{2}>0$. Integrating out the scalars when $m_{\Phi}^{2}>0$ is straightforward, they are simply gapped and cause no change in the Chern-Simons terms so we are left with
\begin{equation}
i\CL_{U}^{II}=N\text{CS}_{k}[c^{\prime}]+N\text{CS}_{k}[\tilde{c}+\tilde{A}_{1}\mathds{1}_{k}]-Nk\text{CS}_{1}[\tilde{A}_{1}]+2Nk\text{CS}_\text{grav}.\label{eq:mdb s+f sector 2}
\end{equation}
We will start by assuming Neumann boundary conditions for the dynamical gauge fields; all the Chern-Simons terms are anomalous in the sense that they result in a non-vanishing current flowing onto the boundary. Fortunately, in Phase II it is straightforward to assign edge modes to compensate for the anomalies.

To start, $N$ right-moving $k$-component fiducial fermions coupled to $c^{\prime}+\tilde{c}+\tilde{A}_{1}\mathds{1}_{k}$ will make $N\text{CS}_{k}[c^{\prime}]+N\text{CS}_{k}[\tilde{c}+\tilde{A}_{1}\mathds{1}_{k}]$ non-anomalous. Note that since we can shift away the $\tilde{A}_{1}$ factor, together these terms are equivalent to a $U(k)_{N}$ Chern-Simons term.

Next, $Nk$ left-moving single-component fiducial fermions will make $Nk\text{CS}_{1}[\tilde{A}_{1}]$ non-anomalous. The newly added $Nk$ left and $N$ right movers respectively generate gravitational Chern-Simons terms $+2Nk\text{CS}_\text{grav}$ and $-2Nk\text{CS}_\text{grav}$, and hence such terms cancel out.

Lastly, we need to make the remaining $+2Nk\text{CS}_\text{grav}$ term non-anomalous.  We thus introduce $Nk$ neutral right-moving single-component fiducial fermions. Moving forward, we note that a positive mass scalar does nothing to the Chern-Simons modes, and so, we will always use the $m_{\Phi}^{2}>0$ (or $m_{\phi}^{2}>0$) regime to determine the fiducial fermions on the scalar end of the dualities.

However, there is one subtlety we have not yet mentioned: introducing the fiducial fermions has given the theory additional symmetries on the boundary. For example, choosing to add $N\times\CL_{ff}^+[c^\prime+\tilde{c}+\tilde{A}_1 \mathds{1}_k]$ introduces a new global $SU(N)$ symmetry on the boundary. We need to be careful with how we are assigning fiducial fermions on both sides of the duality so that their associated global symmetries match. While the global symmetries coming from the fiducial fermions for the background Chern-Simons terms trivially match, the fiducial fermions associated with dynamical gauge fields have no analog on the opposite side of the duality.

Taking care to assign the fiducial fermions for the dynamical gauge fields, recall that Dirichlet boundary conditions not only enhances the global symmetry on the boundary but also eliminates the need to make the dynamical gauge fields non-anomalous. This removes the need to assign fiducial fermions to the dynamical gauge fields for Dirichlet boundary conditions. In fact, the enhanced global symmetry from choosing Dirichlet boundary conditions on one side of the duality exactly match the additional global symmetry from introducing the dynamical fiducial fermions \cite{gaiotto}.

Let us demonstrate this mechanism explicitly in the present example. Table \ref{tab:n and d} summarizes all the fiducial fermions we had to add on both sides of the duality. We just explained this fiducial matter content on the $U$ side and will turn to the $SU$ side momentarily. There are common $Nk$ left-moving fermions (charged under $\tilde{A}_1$) and $Nk$ neutral right-moving fermions on both sides. They give rise to an extra $SU(Nk) \times U(Nk)$ global symmetry on both sides. In addition, there are $N$ fiducial fermions on the $U(k)$ side that have no corresponding fiducial fermions on the $SU(N)$ side.
We can account for the new $SU(N)$ global symmetry from these fiducial fermions by choosing Dirichlet boundary conditions for the dynamical $SU(N)$ gauge field $b$, which will produce a global $SU(N)$ symmetry on the boundary. More generally, choosing Neumann boundary conditions for the gauge fields on one side of the duality is {\emph{only}} consistent with choosing Dirichlet boundary conditions on the other.

To complete the entries in Table \ref{tab:n and d}, let us analyze the fiducial matter content on the $SU$ side. Staying in Phase II and integrating out the $N_{f}$ $N$-component dynamical fermions, we pick up additional Chern-Simons terms, which reduces \eqref{eq:mdb s+f lsu} to
\begin{align}
i\CL_{SU}^{II} =-k\text{CS}_{N}[b^\prime]-kN\text{CS}_{1}[\tilde{A}_{1}].\label{eq:mdb f sector 2B}
\end{align}
At this point if we choose Dirichlet boundary conditions on $b^\prime$, the $-k\text{CS}_{N}[b^\prime]$ term is non-anomalous on its own. First note that we have fermions on this side of the duality and so if we choose appropriate boundary conditions, DWFs can exists and potentially provide the necessary edge modes for \eqref{eq:mdb f sector 2B} to be non-anomalous. However, with a bit of foresight we will choose the fermionic boundary condition which does not allow DWFs to exist in this phase, and hence all of our anomaly cancellation must come from fiducial fermions. This also turns out to be the right choice for matching global symmetries on the boundary.

Specifically, introducing $k$ left-moving $N$-component fiducial fermions coupled to $\tilde{A}_{1}\mathds{1}_{N}$ renders $-kN\text{CS}_{1}[\tilde{A}_{1}]$ non-anomalous. To account for the gravitational Chern-Simons term $-2Nk\text{CS}_\text{grav}$ from the fiducial fermions, we should also introduce $Nk$ right-moving neutral fermions. It is easy to see that the boundary global symmetries match the choice of Neumann boundary conditions on the $U$ side above.

We have completed our first complete dual pair. As pointed out in \cite{gaiotto}, we have seen that the duality-consistent boundary conditions for dynamical gauge fields are Neumann on one side of the duality and Dirichlet on the other with the freedom to assign which side sees which boundary condition.

 There is a second dual pair with the same gauge groups and matter content where we choose Dirichlet boundary conditions on the $U$ side and Neumann boundary conditions on the $SU$ side. We can work out the fiducial fermion content in this pair following the same logic as above.

 Staying in Phase II, on the $SU$ side we now need to assign fiducial fermions to make both terms in \eqref{eq:mdb f sector 2B} non-anomalous. Fortunately this isn't much different from the case considered above, and simply requires the $k$ left-moving $N$-component fiducial fermions be coupled to $b^\prime+\tilde{A}_{1}\mathds{1}_{N}$ instead of just $\tilde{A}_{1}\mathds{1}_{N}$. This renders both $ -k\text{CS}_{N}[b^\prime]-kN\text{CS}_{1}[\tilde{A}_{1}]$ non-anomalous.

Now, consider imposing Dirichlet boundary conditions on the $U$ side in Phase II.  Above, we saw that Neumann boundary conditions required three types of fiducial fermions to render all the terms in \eqref{eq:mdb s+f sector 2} non-anomalous. However, chosing Dirichlet boundary conditions for $c$ means that we no longer need to worry about canceling the anomaly associated with its Chern-Simons term. In this case the anomalies of $N\text{CS}_{k}[\tilde{c}+\tilde{A}_{1}\mathds{1}_{k}]$ and $-Nk\text{CS}_{1}[\tilde{A}_{1}]$ actually cancel, and this means that we only need the fiducial fermions that made the gravitational term non-anomalous.

Having established the fiducial fermion spectrum in Phase II, let's now check that the assignments work to make Phase I non-anomalous as well. For the $SU$ + fermion theory, integrating out the fermions in \eqref{eq:mdb s+f lsu} cancels the $\eta$-invariants, which leaves the Chern-Simons levels unaffected,
\begin{align}
i\CL_{SU}^I & = (N_f -k)\text{CS}_{N}[b^\prime]-(k-N_f)N\text{CS}_{1}[\tilde{A}_1]+N\text{CS}_{N_{f}}[C]+2NN_{f}\text{CS}_\text{grav}.\label{eq:mdb f sector 1}
\end{align}
However, it will be helpful to view $i\CL_{SU}^{I}$ as coming from $i\CL_{SU}^{II}$ in order to show that \eqref{eq:mdb f sector 1} is non-anomalous. Comparing to \eqref{eq:mdb f sector 2B},
\begin{align}\label{eq:f sector 1B}
i\CL_{SU}^{I}=i\CL_{SU}^{II}+N_{f}\text{CS}_{N}[b^\prime]+NN_{f}\text{CS}_{1}[\tilde{A}_{1}]+N\text{CS}_{N_{f}}[C]+2NN_{f}\text{CS}_\text{grav}.
\end{align}

In order to get a non-anomalous theory, we can take advantage of the fact that the fiducial fermions that we have already assigned rendered $i\CL_{SU}^{II}$ non-anomalous. It remains to be shown that the additional Chern-Simons terms in \eqref{eq:f sector 1B} are non-anomalous. Fortunately, we have chosen the boundary condition on the dynamical fermion such that we allow the DWFs to live for $m_{\psi}>0$. From \eqref{eq:su_prime}, the dynamical fermions couple to $b^\prime\mathds{1}_{N_f} +C\mathds{1}_{N}+ \tilde{A}_1\mathds{1}_{NN_f}$, and hence the DWFs are exactly the edge modes needed to cancel the residual anomalies of \eqref{eq:f sector 1B}.

The cancellation of the edge modes happens analogously to the cancellation of the Chern-Simons terms. In the end, we have $k-N_f$ left-moving $N$-component fiducial fermions coupled to $b^\prime+\tilde{A}_{1}\mathds{1}_{N}$,  and $N$ right-moving $N_f$-component fiducial fermions coupled to $C$ to cancel the background $SU(N_f)$ and gravitational Chern-Simons terms.

To complete the duality in Phase I, we need to consider the scalar side in the Higgs regime ($m^2_\Phi <0$). Following \cite{Jensen:2017bjo, Benini:2017aed}, we will assume that the $N_{f}$ scalars maximally Higgs the $U(k)$.  The breaking pattern is then $U(k)_{-N}\to U\left(k-N_{f}\right)_{-N}\times SU\left(N_{f}\right)_{-N}$; resulting in a Lagrangian
\begin{align}
i\CL_{U}^{I} & =N\left(\text{CS}_{k-N_{f}}[c^{\prime}]+\text{CS}_{k-N_{f}}[\tilde{c}+\tilde{A}_{1}\mathds{1}_{k-N_{f}}]+\text{CS}_{N_{f}}[C] -(k-N_{f})\text{CS}_{1}[\tilde{A}_{1}]+2k\text{CS}_\text{grav}\right).\label{eq:mdb s+f 1}
\end{align}
Since there are fiducial fermions which couple to $U(k)$, the spontaneous breaking separates each of the $k$-component fiducial fermions into broken and unbroken parts, namely
\begin{align}\label{eq:breaking_U}
N\times\CL_{ff}^+[c^\prime+\tilde{c}+\tilde{A}_1\mathds{1}_{k}]\to \begin{cases}
N\times\CL_{ff}^+[c^\prime+\tilde{c}+\tilde{A}_1\mathds{1}_{k-N_{f}}] & \text{(unbroken)}\\
N\times\CL_{ff}^+[C+\tilde{A}_1\mathds{1}_{N_f}] & \text{(broken)}
\end{cases}.
\end{align}
Note that the $\tilde{A}_{1}$ part of the $N$ $N_f$-component fiducial fermions from the broken sector combines with the opposite chirality $Nk$ modes coupled to $\tilde{A}_{1}$; leaving a total of $N(k-N_{f})$. The number of gravitational Chern-Simons terms is unchanged -- we still have the same net number of modes. A straightforward check shows these edge modes render \eqref{eq:mdb s+f 1} non-anomalous.

Comparing the boundary spectra for mass deformations of the single species non-Abelian duality, we can match the degrees of freedom in kind.  Thus, we see that
\begin{align}
\CL_{U}^{I} & \leftrightarrow\CL_{SU}^{I}\\
\CL_{U}^{II} & \leftrightarrow\CL_{SU}^{II}
\end{align}
indicating a consistent duality in the bulk.  We outline both instances of duality consistent boundary conditions and the additional edge modes in Table \ref{tab:n and d}.

\begin{table}
\begin{centering}
\begin{tabular}{|>{\centering}p{5cm}|>{\centering}p{5cm}|>{\centering}p{5cm}|}
\hline
 & \textbf{$SU(N)$ + fermions} & \textbf{$U(k)$ + scalars}\tabularnewline
\hline

\textbf{Boundary Conditions} & $\psi^{-}_{\alpha I}|_{\partial}=0$

$b^\prime$: Dirichlet & $\Phi_{\rho I}$: Dirichlet

$c$: Neumann \tabularnewline

\hline

\textbf{Additional Edge Modes}& {\bf{---}{---}{---} }& $N\times\CL_{ff}^{+}[c^{\prime}+\tilde{c}+\tilde{A}_{1}\mathds{1}_{k}]$ \tabularnewline
\cline{2-3}
 &  $Nk\times\CL_{ff}^{-}[\tilde{A}_{1}]$& $Nk\times\CL_{ff}^{-}[\tilde{A}_{1}]$\tabularnewline
\cline{2-3}
 & $Nk\times\CL_{ff}^{+}[0]$  & $Nk\times\CL_{ff}^{+}[0]$\tabularnewline

\hline
\hline
 & \textbf{$SU(N)$ + fermions} & \textbf{$U(k)$ + scalars}\tabularnewline
\hline

\textbf{Boundary Conditions} & $\psi^{-}_{\alpha I}|_{\partial}=0$

$b^\prime$: Neumann & $\Phi_{\rho I}$: Neumann

$c$: Dirichlet \tabularnewline

\hline

\textbf{Additional Edge Modes}&  $k\times\CL_{ff}^{-}[b^\prime+\tilde{A}_{1}\mathds{1}_{N}]$& {\bf{---}{---}{---}}\tabularnewline
\cline{2-3}
 & $Nk\times\CL_{ff}^{+}[0]$  & $Nk\times\CL_{ff}^{+}[0]$\tabularnewline

\hline
\end{tabular}
\par\end{centering}
\caption{The top (bottom) table counts the additional edge modes when choosing Neumann and Dirichlet boundary conditions on the dynamical gauge fields in $U$ ($SU$) and $SU$ ($U$) side respectively when $N_s=0$.  \label{tab:n and d}}
\end{table}

The last remaining question we have to address is how to identify boundary conditions for the scalar fields. Following a similar procedure used in \cite{Aitken:2017nfd}, let us reinterpret the effect of anomaly inflow when we choose Neumann boundary conditions on the $U$ side. Alone, a Chern-Simons term is anomalous on the boundary due to a non-trivial current divergence. Since the associated current is not conserved, we can think of this as meaning the $U(1)_m$ symmetry is broken on the boundary. When we introduce edge modes on the boundary, there is a compensating term for the current flowing onto the boundary. In other words, if we identify the $U(1)$ axial symmetry on the boundary with the $U(1)_m$ symmetry in the bulk, we have a restored $U(1)$ symmetry everywhere. This is consistent with the $SU$ side of the theory where there is no anomalous term and thus the $U(1)_b$ symmetry exists everywhere.

If we choose Neumann boundary conditions for $c$ on the $U$ side of the duality, this amounts to the constraint that $F_{yi}|_{\partial}=0$, with $F$ the field strength of $c$. Since the flux current is $j^{\mu}_{\text{flux}}\sim \epsilon^{\mu\nu\rho}F_{\nu\rho}$, Neumann boundary conditions automatically imply any flux current on the boundary must vanish. This is consistent with the $U(1)$ boundary symmetry being provided by the edge modes, rather than the flux current. The Neumann boundary condition on the gauge fields is also inconsistent with having a scalar current on the boundary since such a current is charged under the dynamical gauge field. Additionally, recall that the bulk equations of motion relate bulk flux and matter currents; schematically, $j^\mu_{\text{matter}}\sim j_{\text{flux}}^{\mu}$. Although such equations do not apply on the boundary, allowing for scalar current to flow on the boundary would be inconsistent with the continuity of the current and also have no compensating current on the $SU$ side. Therefore, we choose Dirichlet boundary conditions for the scalar which kills off the scalar current on the boundary.

Now consider Dirichlet boundary conditions for $c$. Although $c_i|_{\partial}=0$, this does not necessarily imply $F_{yi}|_{\partial}=0$ since $\partial_i c_y|_{\partial} \ne 0$ (although it does imply $F_{ij}|_{\partial}=0$). By the same reasoning above, this means we can have a nonzero flux current on the boundary. Such boundary conditions are consistent with there being matter charged under the dynamical gauge field on the boundary. The only boundary condition that is consistent with this is Neumann boundary conditions for the scalar. Again via the identification of global symmetry currents, we see that this is consistent with choosing Neumann boundary conditions for $b_\mu^\prime$ since we have a nonzero edge modes coupling to $\tilde{A}_1$ on the $SU$ end now.

\subsubsection*{Abelian Reduction}

Let's apply a consistency check on our new non-Abelian prescription. We will take the limit $N_{f}=N=k=1$, and choose Neumann boundary conditions for $c$ on the $U$ side to compare to the boundary analysis of the Abelian dualities \cite{Aitken:2017nfd}.  Affecting this limit in (\ref{eq:mdb s+f lsu}) and (\ref{eq:mdb u+s lu}) gives
\begin{align}
\CL_{SU} & =i\bar{\psi}\Dslash_{\tilde{A}_{1}}\psi-i\left(2\text{CS}_\text{grav}\right)\\
\CL_{U} & =\left|D_{\tilde{c}}\Phi\right|^{2}+\alpha_\varphi|\Phi|^4-i\left(\text{CS}_{1}[\tilde{c}+\tilde{A}_1]+\text{CS}_1[\tilde{c}]+2\text{CS}_\text{grav}\right),
\end{align}
which is similar to the Abelian ``scalar + flux = fermion'' considered in \cite{Aitken:2017nfd}, up to the additional $2\text{CS}_\text{grav}$ terms.

Now, taking the Abelian limit of the tallied boundary modes in Table \ref{tab:n and d}, we find that one fiducial fermion is needed on $U$ side to be coupled to $\tilde{c}+\tilde{A}_1$ and, on both sides of the duality, we need one left-mover coupled to $\tilde{A}_1$ and a neutral right-mover.

Due to certain subtleties with the non-Abelian case, our convention has changed slightly as compared to \cite{Aitken:2017nfd} where the opposite boundary conditions on the dynamical fermions were chosen and gravitational Chern-Simons terms were absent. Without gravitational Chern-Simons terms present we do not need the right-moving neutral fiducial fermions on both sides of the duality. Choosing opposite boundary conditions on the dynamical fermions makes the $m_\Psi<0$ regime consistent via a fiducial fermion rather than a dynamical fermion. This is why in the present analysis we find an additional left-moving fiducial fermion coupled to $\tilde{A}_1$ on the fermion side of the duality. Choosing Dirichlet boundary conditions on the scalar was also found for a similar reason. Thus, the number of edge modes is consistent modulo conventions.

Notice that the fermions couple to the background $U(1)_m$ spin$_c$ connection, $\tilde{A}_1$. The analysis in \cite{Aitken:2017nfd} requires that in order for the ``scalar+flux = fermion'' duality to be consistent in the presence of a boundary $\tilde{A}_1$ must be a spin$_c$ connection and not an ordinary $U(1)$. Meanwhile, $\tilde{c}$ was required to be an ordinary connection. Indeed, both of these requirements are consistent the Abelian limit.

\subsection{Non-Abelian $U+\text{fermions}\leftrightarrow SU+\text{scalars}$}
\label{sec:nab 2}

Now let us consider the other type of single species non-Abelian duality in \cite{Aharony:2015mjs} -- rather its time reversed version -- by setting $N_{f}=0$ such that (\ref{eq:level-rank}) reads
\begin{equation}
SU(N)_{-k}\,\text{with \ensuremath{N_{s}} \ensuremath{\phi}}\qquad\leftrightarrow\qquad U(k)_{N-\frac{N_{s}}{2}}\,\text{with \ensuremath{N_{s}} \ensuremath{\Psi}}.
\end{equation}
with the mass identification $m_{\phi}^{2}\leftrightarrow m_{\Psi}$. In this case the flavor bound is given by $N_{s}\leq N$ \cite{Benini:2017aed, Jensen:2017bjo}. The explicit Lagrangians for the theories on each side of the duality are given by
\begin{align}
\CL_{SU} =&\left|D_{b+B}\phi\right|^{2}+\alpha_\varphi |\phi|^4-i\left(-k\text{CS}_{N}[b]+\text{BF}[f;\text{Tr}_{N}\left(b\right)-N\tilde{B}]\right)\nonumber \\
& -i\left(Nk\text{CS}_1[\tilde{B}]-Nk\text{CS}_1[\tilde{A}_1] \right) \label{eq:mdb s lsu},\\
\CL_{U} =&i\bar{\Psi}\Dslash_{c-\tilde{A}_2+B}\Psi-i\left(N\text{CS}_{k}[c]+N\text{BF}[\text{Tr}_{k}\left(c\right);\tilde{A}_{1}]+2Nk\text{CS}_\text{grav}\right)\nonumber \\
 =&i\bar{\Psi}\Dslash_{c^\prime+\tilde{a}+B}\Psi-i\left(N\text{CS}_{k}[c^{\prime}]+N\text{CS}_{k}[\tilde{a}+\tilde{B}\mathds{1}_{k}]-Nk\text{CS}_{1}[\tilde{A}_{1}]+2Nk\text{CS}_\text{grav}\right).\label{eq:mdb f+f lu}
\end{align}
Setting $N_{f}=0$ has eliminated one of the gravitational Chern-Simons terms, and in the last line of \eqref{eq:mdb f+f lu} we defined the ordinary connection $\tilde{B}=\tilde{A}_1+\tilde{A}_2$ and spin$_c$ connection $\tilde{a} = \tilde{c}-\tilde{A}_2 \mathds{1}_k$. Note that $\tilde{B}$ is now the background gauge field associated with the global $U(1)_{m,b}$ symmetry. We have also used
\begin{align}
\text{BF}[\tilde{A}_{1};\tilde{A}_{2}]+\text{CS}_1[\tilde{A}_{2}] = \text{CS}_1[\tilde{B}]-\text{CS}_1[\tilde{A}_1].
\end{align}
 For this dual pair, mass deformations correspond to Phase II ($m_{\phi}^{2}>0$ $m_{\Psi}>0$ ) and Phase III ($m_{\phi}^{2}<0$ $m_{\Psi}<0$) -- see Fig. \ref{fig:Various-phases-of master mdb}.  As with the $N_s=0$ case, we can find the fiducial fermion spectrum by looking at Phase II.

As with the last duality, we will find the boundary symmetries to be consistent only if we choose Neumann and Dirichlet boundary conditions for the dynamical gauge fields on opposite sides of the duality. Nevertheless, we will first proceed with the analysis for Neumann boundary conditions on both sides of the duality; generalizing to Dirichlet is straightforward. For the $SU$ + scalar theory in \eqref{eq:mdb s lsu} with Neumann boundary conditions for the dynamical gauge fields, integrating out the Lagrange multiplier gives
\begin{align}
i\CL_{SU}^{II}=-k\text{CS}_{N}[b^\prime]-kN\text{CS}_{1}[\tilde{A}_{1}].
\end{align}
$k$ left-moving $N$-component fiducial fermions coupled to $b^\prime+\tilde{A}_{1}\mathds{1}_{N}$ compensate for the anomalies generated by $-k\text{CS}_{N}[b^\prime]-kN\text{CS}_{1}[\tilde{A}_{1}]$. We also need $Nk$ right-moving neutral fiducial fermions to cancel the gravitational term.

The $U$ side of the duality is also easy to analyze with Neumann boundary conditions.  Despite the new definitions of $\tilde{B}$ and $\tilde{a}$, the anomaly spectrum of (\ref{eq:mdb f+f lu}) is identical to that of (\ref{eq:mdb s+f sector 2}). We can choose exactly the same fiducial fermions for the $U$ + fermion side of the duality that we did for $U$ + scalar with Neumann conditions in Table \ref{tab:n and d}.

Having quickly read off the fiducial fermions in Phase II, we should check that the assignment holds for Phase III.  In Phase III for the $U$ + fermion theory $m_\Psi <0$, and so, integrating out the dynamical fermions shifts the Chern-Simons levels relative to their Phase II values:
\begin{equation}
i\CL_{U}^{II} \rightarrow i\CL_U^{II} -N_{s}\text{CS}_{k}[c]-N_sk\text{CS}_1[\tilde{A}_2]-k\text{CS}_{N_{s}}[B]-2N_{s}k\text{CS}_\text{grav}.\label{eq:mdb f+f 1}
\end{equation}
Using the first line of (\ref{eq:mdb f+f lu}), the Lagrangian for the $U$ + fermion theory becomes
\begin{align}
i\CL_{U}^{III} & =\left(N-N_{s}\right)\text{CS}_{k}[c]-k\text{CS}_{N_{s}}[B]+N\text{BF}[\text{Tr}_{k}\left(c\right);\tilde{A}_{1}\mathds{1}_k]-N_sk\text{CS}_1[\tilde{A}_2]+2k\left(N-N_{s}\right)\text{CS}_\text{grav}.\label{eq:mdb f+f sector3A}
\end{align}
Rewriting the BF term as a sum of Chern-Simons terms, we find
\begin{align}\begin{split}
i\CL_U^{III}& =\left(N-N_{s}\right)\left(\text{CS}_{k}[c^{\prime}]+\text{CS}_{k}\left[\tilde{c}+\frac{N}{N-N_{s}}\tilde{A}_{1}\mathds{1}_k\right]+2k\text{CS}_\text{grav}\right)\\&\quad-k\left(\text{CS}_{N_{s}}[B]+N\text{CS}_{1}[\tilde{A}_{1}]+\frac{NN_s}{\left(N-N_{s}\right)}\text{CS}_{1}[\tilde{B}]\right).\label{eq:mdb f+f sector3B}
\end{split}\end{align}
So long as we choose the boundary condition such that dynamical DWFs are allowed for $m_{\Psi}<0$, the $U$ + fermion theory in Phase III non-anomalous theory. This follows for the same reason we saw in Phase I of the $SU$ side in Sec. \ref{sec:nab 1}: from \eqref{eq:mdb f+f 1} $i\CL_U^{II}$ is already non-anomalous due to the fiducial fermions and the dynamical DWF provides the rest of the edge modes to render the whole expression non-anomalous. Thus, the fiducial fermion assignment for Phase II works in Phase III, and the $U$ + fermion theory is non-anomalous. While it may be hard to see that \eqref{eq:mdb f+f sector3B} is non-anomalous, the cancelling of the edge modes can be seen directly from the cancellation of the Chern-Simons terms. Finally, note when one expands out $\tilde{B}$ in \eqref{eq:mdb f+f sector3B} this reproduces the stated background terms of \eqref{eq:phaseIII bkg}, as it should.

The $SU$ + scalar theory in Phase III ($m_\phi^2 <0$) is complicated slightly due to the Lagrange multiplier -- which changes $SU(N)\to U(N)\times U(1)$ and makes the breaking pattern clearer. We do not want to treat the BF terms containing the Lagrange multiplier as additional Chern-Simons terms. We will be more concerned with analyzing the behavior of the edge modes after the breaking has occurred as above on the $U$ side.

After spontaneously breaking $U(N)\rightarrow U(N-N_s)\times SU(N_s)$, $N-N_{s}$ scalars remain coupled to $b^{\prime}+y\tilde{B}\mathds{1}_{N-N_{s}}$. The $N_{s}$-components corresponding to the broken part of the gauge symmetry have no coupling to any part of $b^{\prime}$ but do couple to the $SU\left(N_{s}\right)$ flavor symmetry. The factor $y$ is a rescaling of the Abelian coupling implemented by the Lagrange multiplier that is novel to this theory.  Explicitly, the coupling of the $N-N_{s}$ modes now becomes
\begin{equation}
b^{\prime}+y\tilde{B}\mathds{1}_{N}\to b^{\prime}+\frac{\sqrt{N}}{N-N_{s}}\tilde{B}\mathds{1}_{N-N_{s}}.
\end{equation}
Thus, when one integrates out the $k$ fiducial fermions, they give
\begin{equation}
i\CL_{SU}^{III}\supset-k\left(\text{CS}_{N-N_{s}}[b^{\prime}]+\frac{N}{N-N_{s}}\text{CS}_{1}[\tilde{B}]\right),
\end{equation}
which will combine with the existing background terms to reproduces the Abelian factor in \eqref{eq:mdb f+f sector3B}.

Let us choose Neumann boundary conditions for $b^\prime$.  The dividing of the fiducial fermion that we would assign occurs analogously to the breaking of the Chern-Simons terms:
\begin{align}\label{eq:breaking_SU}
k\times\CL_{ff}^+[b^\prime+\tilde{B}]\to \begin{cases}
k\times\CL_{ff}^+\left[b^{\prime}+\frac{\sqrt{N}}{N-N_{s}}\tilde{B}\mathds{1}_{N-N_{s}}\right] & \text{(unbroken)}\\
k\times\CL_{ff}^+[B] & \text{(broken)}
\end{cases}.
\end{align}
There are still $Nk$ total fermion components; $N_{s}k$ of which couple only to the flavor symmetry. Thus, we still have the same number of gravitational Chern-Simons terms as in Phase II. The full Lagrangian for the $SU$ side of Phase III is then
\begin{equation}\label{eq:nab2 1}
i\CL_{SU}^{III}=-k\text{CS}_{N-N_{s}}[b^{\prime}]-k\left(\text{CS}_{N_{s}}[B]-N\text{CS}_{1}[\tilde{A}_{1}]+\frac{NN_s}{N-N_{s}}\text{CS}_{1}[\tilde{B}]\right),
\end{equation}
which is rendered non-anomalous by the edge modes from the fiducial fermions as assigned in Phase II.

Thus far, we have only considered Neumann boundary conditions for the dynamical gauge fields. To generalize these results to the Dirichlet case is straightforward: simply remove the coupling of the fiducial fermion to the dynamical field whose Chern-Simons terms is no longer anomalous on the boundary. Table \ref{tab:n vs d 2} summarizes our results for this duality. Note once again a nice cancellation between anomalous terms occurs on the $U$ side with Dirichlet boundary conditions.

Finally, consider the boundary conditions on the scalar fields. Again, we use fact that Neumann boundary conditions imply any flux current on the boundary must vanish and that the variation of $\tilde{A}_1$ relates the scalar matter current on the $SU$ side to the flux current on the $U$ side. Since there can be no flux current on the boundary, there can be no scalar current on the boundary as well. Hence we must choose Dirichlet boundary conditions in this case, $\phi_{\alpha M}|_{\partial}=0$.

As we argued earlier, for Dirichlet boundary conditions on $c$ we can have a nonzero flux current on the boundary. Again using the identification of global symmetry currents we can also have a nonzero scalar current on the $SU$ side of the duality. Thus, we must choose scalar boundary conditions which allow for a nonzero boundary current, which means Neumann.

\begin{table}
\begin{centering}
\begin{tabular}{|>{\centering}p{5cm}|>{\centering}p{5cm}|>{\centering}p{5cm}|}
\hline
 & \textbf{$SU(N)$ + scalars} & \textbf{$U(k)$ + fermions}\tabularnewline
\hline

\textbf{Boundary Conditions} & $\phi_{\alpha M}$: Neumann

$b^\prime$: Neumann & $\Psi^{+}_{\rho M}|_{\partial}=0$

$c$: Dirichlet \tabularnewline

\hline

\textbf{Additional Edge Modes} & $k\times\CL_{ff}^{-}[b^\prime+\tilde{A}_{1}\mathds{1}_{N}]$  & {\bf{---}{---}{---}} \tabularnewline
\cline{2-3}
 & $Nk\times\CL_{ff}^{+}[0]$  & $Nk\times\CL_{ff}^{+}[0]$\tabularnewline

\hline
\hline
 & \textbf{$SU(N)$ + scalars} & \textbf{$U(k)$ + fermions}\tabularnewline
\hline

\textbf{Boundary Conditions} & $\phi_{\alpha M}$: Dirichlet

$b^\prime$: Dirichlet & $\Psi^{+}_{\rho M}|_{\partial}=0$

$c$: Neumann \tabularnewline

\hline

\textbf{Additional Edge Modes} &{\bf{---}{---}{---}} & $N\times\CL_{ff}^{+}[c^{\prime}+\tilde{a}+\tilde{B}\mathds{1}_{k}]$

$=N\times\CL_{ff}^{+}[c^{\prime}+\tilde{c}+\tilde{A}_{1}\mathds{1}_{k}]$\tabularnewline
\cline{2-3}
 & $Nk\times\CL_{ff}^{-}[\tilde{A}_{1}]$  & $Nk\times\CL_{ff}^{-}[\tilde{A}_{1}]$\tabularnewline
\cline{2-3}
 & $Nk\times\CL_{ff}^{+}[0]$  & $Nk\times\CL_{ff}^{+}[0]$\tabularnewline

\hline
\end{tabular}
\par\end{centering}
\caption{The top (bottom) table counts the additional edge modes when choosing Neumann and Dirichlet boundary conditions on the dynamical gauge fields in $SU$ ($U$) and $U$ ($SU$) side respectively when $N_f=0$.\label{tab:n vs d 2}}
\end{table}

\subsubsection*{Abelian Reduction}

Finally, let us check that this is consistent in the Abelian limit by setting $N=k=N_{s}=1$, choosing Neumann boundary conditions for $\tilde{c}$, and moving all background terms to the fermion side.  Affecting this limit, we find
\begin{align}
\CL_{SU} & =\left|D_{\tilde{B}}\phi\right|^{2}+\alpha_\varphi |\phi|^4\\
\CL_{U} & =i\bar{\Psi}\Dslash_{\;\tilde{a}}\Psi-i\left(\frac{1}{4\pi}\text{CS}_{1}[\tilde{a}+\tilde{B}]+2\text{CS}_\text{grav}\right)
\end{align}
where we have canceled the two $-\text{CS}_{1}[\tilde{A}_{1}]$ terms. This expression should be equivalent to the \emph{time-reversed} fermion, but with the understanding that in \cite{Aitken:2017nfd} the time-reversed fermion came with an opposite sign Pauli-Villars regulator as well; our conventions for the $\eta$-invariant are different here.  Accounting for this difference of convention, we pick up an overall shift by $-\text{CS}_{1}[\tilde{a}]-2\text{CS}_\text{grav}$ on the fermionic side of the duality and change the fermionic boundary condition. We end up with the dual theories being given by
\begin{align}
\CL_{SU} & =\left|D_{\tilde{B}}\phi\right|^{2}+\alpha_\varphi|\phi|^4,\\
\CL_{U} & =i\bar{\Psi}\Dslash_{\;\tilde{a}}\Psi-i\left(\text{CS}_{1}[\tilde{a}+\tilde{B}]-\text{CS}_{1}[\tilde{a}]\right).
\end{align}
Per our fiducial fermion choices shown in Table \ref{tab:n vs d 2}, we should have a single right-moving fiducial fermion coupled to $\tilde{a}+\tilde{B}$. Note the fermions associated to $\tilde{B}$ and neutral fiducial fermions on both ends of the duality cancel one another out.

Once more we see a nice consistency with our previous analysis: $\tilde{a}=\tilde{c}-\tilde{A}_2$ is a spin$_c$ connection, and the background field  $\tilde{B}=\tilde{A}_1+\tilde{A}_2$ is an ordinary $U(1)$ connection.  Thus, we can start from the master bosonization duality, demand that a subset of Abelian factors be either ordinary or spin$_c$ connections, and consistently arrive at {\emph{both}} known Abelian bosonization dualities with the correct coupling of gauge fields to matter.  This is also consistent with the process of promoting background fields to dynamical and coupling to new background fields followed by integrating out the old dynamical fields \cite{Karch:2016sxi, Seiberg:2016gmd}.

\subsection{Discussion}

Before turning back to the master bosonization duality, let us take stock of how the phases and edge modes changed when we moved to negative mass deformations for the fermions and scalars:
\begin{itemize}
\item \textbf{Fermion Deformations: }Given our choice of fermionic boundary conditions, the dynamical DWFs only existed when $m_{\psi}>0$ or $m_{\Psi}<0$. In the corresponding $m_\psi<0$ and $m_\Psi>0$ phases, we found that the additional Chern-Simons terms were rendered non-anomalous by the dynamical DWFs. Since the $m_{\psi}>0$ and $m_{\Psi}<0$ phases were non-anomalous due to the fiducial fermions, the resulting theory was non-anomalous. Furthermore, the same mechanism that rendered the Chern-Simons terms non-anomalous can be used to argue that -- despite some simplified forms of the theories appearing to have extraneous edge modes -- that edge modes are cancelled.
\item \textbf{Scalar Deformations:} In the spontaneously broken phase, $m_{\phi}^{2}<0$ or $m_{\Phi}^{2}<0$, the dynamical gauge groups are split up into smaller dynamical groups and gave rise to new non-Abelian flavor symmetries. Additionally for $SU$ + scalars, the background Abelian coupling was rescaled. The couplings of the fiducial fermions were changed according to the breaking pattern for the Chern-Simons terms. The fiducial fermions then split into parts, which couple to the broken and unbroken parts of the gauge group. The remaining dynamical and new flavor Chern-Simons terms are rendered non-anomalous by this set of fiducial fermions.
\end{itemize}
Although the master duality is slightly more complicated due to two independent mass deformations, we will see that the same mechanisms that lead to non-anomalous theories in both phases of the single-species non-Abelian cases completely generalize. Since the fiducial fermions make the positive mass phase non-anomalous and the fiducial/dynamical fermions -- including the singlet -- continue to work after Higgsing or integrating out negative mass fermions, all five phases of the master duality continue to be non-anomalous.

\section{Master Duality with Boundaries}\label{sec:master-boundaries}

Now that we have firmly established how to derive the correct set of boundary conditions and assignments of fiducial fermions in order to render boundary theories non-anomalous in the single-species non-Abelian dualities, we can analyze the two-species master bosonization duality.  Having made the assignments in the common Phase II region, the fiducial fermions of the two single-species non-Abelian cases considered are consistent with one another --  see Tables \ref{tab:n and d} and \ref{tab:n vs d 2}.  We can then combine the two prescriptions and check their compatibility across all five mass deformed regions in Fig.~\ref{fig:Various-phases-of master mdb}.

\begin{table}
\begin{centering}
\begin{tabular}{|>{\centering}p{5cm}|>{\centering}p{5cm}|>{\centering}p{5cm}|}
\hline
 & \textbf{$SU(N)$ Side} & \textbf{$U(k)$ Side}\tabularnewline
\hline

\textbf{Boundary Conditions} & $\psi^{-}_{\alpha I}|_{\partial}=0$

$\phi_{\alpha M}$: Neumann

$b^\prime$: Neumann & $\Psi^{+}_{\rho M}|_{\partial}=0$

$\Phi_{\rho I}$: Neumann

$c$: Dirichlet \tabularnewline

\hline

\textbf{Additional Edge Modes} &  $k\times\CL_{ff}^{-}[b^\prime+\tilde{A}_{1}\mathds{1}_{N}]$& {\bf{---}{---}{---}}\tabularnewline
\cline{2-3}
 & $Nk\times\CL_{ff}^{+}[0]$  & $Nk\times\CL_{ff}^{+}[0]$\tabularnewline

\hline
\hline
 & \textbf{$SU(N)$ Side} & \textbf{$U(k)$ Side}\tabularnewline
\hline

\textbf{Boundary Conditions} & $\psi^{-}_{\alpha I}|_{\partial}=0$

$\phi_{\alpha M}$: Dirichlet

$b^\prime$: Dirichlet & $\Psi^{+}_{\rho M}|_{\partial}=0$

$\Phi_{\rho I}$: Dirichlet

$c$: Neumann\tabularnewline

\hline

\textbf{Additional Edge Modes} &{\bf{---}{---}{---}} & $N\times\CL_{ff}^{+}[c^{\prime}+\tilde{c}+\tilde{A}_{1}\mathds{1}_{k}]$  \tabularnewline
\cline{2-3}
 &  $Nk\times\CL_{ff}^{-}[\tilde{A}_{1}]$ & $Nk\times\CL_{ff}^{-}[\tilde{A}_{1}]$\tabularnewline
\cline{2-3}
 & $Nk\times\CL_{ff}^{+}[0]$  & $Nk\times\CL_{ff}^{+}[0]$\tabularnewline

\hline
\end{tabular}
\par\end{centering}
\caption{The top (bottom) table counts the additional edge modes when choosing Neumann and Dirichlet boundary conditions on the dynamical gauge fields in $SU$ ($U$) and $U$ ($SU$) side respectively when $N_f\neq0$ and $N_s\neq 0$. \label{tab:n and d master}}
\end{table}

We will analyze the phases on the $U$ and $SU$ sides roughly in order of increasing difficulty. The discussion will be kept brief for phases where cancellation is a straightforward generalization of what we have already observed in the single-species non-Abelian cases of Sec. \ref{sec:single-species-boundaries}.  In the following analysis, we are interested in the assignments that render the theories non-anomalous, and so we will assume Neumann conditions on the dynamical gauge fields throughout.  Although Neumann boundary conditions on both dynamical gauge fields does not yield a consistent duality, generalization to Dirichlet boundary conditions for one of the dynamical gauge fields is straightforward, see Sec. \ref{sec:nab 2}.

\subsection*{Phase II}

This phase corresponds to $m_{\psi}<0$ and $m_{\phi}^{2}>0$ on the $SU$ side and $m_{\Psi}>0$ and $m_{\Phi}^{2}>0$ on the $U$ side. Starting from \eqref{eq:mdb lsu} and \eqref{eq:mdb lu}, after integrating out all of the matter fields, we find that
\begin{align}
i\CL_{SU}^{II} =-k\text{CS}_{N}[b]+\text{BF}[f;\text{Tr}_{N}\left(b\right)-N\tilde{A}_{1}-N\tilde{A}_{2}]+Nk\text{BF}[\tilde{A}_{1};\tilde{A}_{2}]+Nk\text{CS}_{1}[\tilde{A}_{2}],\label{eq:mdb phaseII lsu}
\end{align}
After some simplification, \eqref{eq:mdb lsu} and \eqref{eq:mdb lu} reduce to
\begin{align}
i\CL_{SU}^{II} & =-k\text{CS}_{N}[b^\prime]-Nk\text{CS}_{1}[\tilde{A}_{1}],\\
i\CL_{U}^{II} & =N\text{CS}_{k}[c^{\prime}]+N\text{CS}_{k}[\tilde{c}+\tilde{A}_{1}\mathds{1}_{k}]-Nk\text{CS}_{1}[\tilde{A}_{1}]+2Nk\text{CS}_\text{grav}.
\end{align}
Since this was the phase of the duality where we chose all of our fiducial fermions such that the theory was non-anomalous, no further analysis is needed, and the assignments are listed in Table \ref{tab:n and d master}

\subsection*{Phase I}

This phase corresponds to $m_{\psi}>0$ and $m_{\phi}^{2}>0$ on the $SU$ side and $m_{\Psi}>0$ and $m_{\Phi}^{2}<0$ on the $U$ side.

\subsubsection*{$U$ Side}

For $m_{\Psi}>0$, the Chern-Simons levels are unaffected when integrating out the fermions. However because  $m_{\Phi}^{2}<0$, the theory is in a spontaneously broken phase
\begin{align}
\hspace{-0.15cm}i\CL_{U}^{I} & =N\left(\text{CS}_{k-N_{f}}[c^{\prime}]+\text{CS}_{N_{f}}[C]+\text{CS}_{k-N_{f}}[\tilde{c}+\tilde{A}_{1}\mathds{1}_{k}]-\left(k-N_{f}\right)\text{CS}_{1}[\tilde{A}_{1}]+2k\text{CS}_\text{grav}\right)
\end{align}
As with the single-species non-Abelian case, the edge modes automatically split up to make the new Chern-Simons modes non-anomalous. The original $N$ right-moving $k$-component fiducial fermions break in a manner completely analogous to \eqref{eq:breaking_U}. The modes coupling to the unbroken $U(k-N_f)$ render $N(\text{CS}_{k-N_{f}}[c^{\prime}] +\text{CS}_{k-N_{f}}[\tilde{c}+\tilde{A}_{1}\mathds{1}_{k}])$ non-anomalous. The parts of the $N_f$-component modes coupling to $\tilde{A}_{1}$ can cancel with the fiducial fermions of opposite chirality which only couple to $\tilde{A}_{1}$, leaving only the $C$ coupling. Hence, the $N\text{CS}_{N_{f}}[C]$ and $-N\left(k-N_{f}\right)\text{CS}_{1}[\tilde{A}_{1}]$ terms are also non-anomalous. Since the number of fiducial fermions hasn't changed at all, the gravitational Chern-Simons term is also still non-anomalous.

\subsubsection*{$SU$ Side}

On this side of the duality, we have $m_{\psi}>0$ and $m_{\phi}^{2}>0$. Neither the scalar nor the fermion change the Chern-Simons terms when integrated out. Note that we have chosen the boundary conditions on the dynamical fermion such that we let the $\psi$ DWFs exist in this phase.

The fact that the theory is non-anomalous, however, should be evident if we rewrite $\CL_{SU}^{I}$ in terms of $\CL_{SU}^{II}$,
\begin{align} \label{eq:mdb phaseI lsu}
i\CL_{SU}^{I}&=i\CL_{SU}^{II}+N_f\text{CS}_{N}[b^\prime]+N\text{CS}_{N_f}[C]+NN_f\text{CS}_{1}[\tilde{A}_1]+2NN_f \text{CS}_\text{grav} \nonumber\\
&=-(k-N_f)\text{CS}_{N}[b^\prime]+N\text{CS}_{N_f}[C]-N(k-N_f)\text{CS}_{1}[\tilde{A}_1]+2NN_f \text{CS}_\text{grav}.
\end{align}
We already have assigned the fiducial fermions so that the $i\CL_{SU}^{II}$ is non-anomalous. Provided that the dynamical DWFs are enough to make the new Chern-Simons terms non-anomalous, the entire Lagrangian in \eqref{eq:mdb phaseI lsu} will be non-anomalous. Since the dynamical fermions couple to $b^\prime+C+\tilde{A}_1$ this is indeed the case. The DWFs cancel with the existing $N_f$ fiducial fermion edge modes, making \eqref{eq:mdb phaseI lsu} non-anomalous.

\subsection*{Phase III}

This phase corresponds to $m_{\psi}<0$ and $m_{\phi}^{2}<0$ on the $SU$ side and $m_{\Psi}<0$ and $m_{\Phi}^{2}>0$ on the $U$ side.

\subsubsection*{$SU$ Side}

The gauge group is spontaneously broken in this phase, but since $m_{\psi}<0$ we have no additional shift of Chern-Simons terms due to integrating out the fermion, relative to our fiducial fermion assignments of Phase II. Spontaneously breaking $SU(N)$ causes the Lagrangian to be modified to
\begin{align}
i\CL_{SU}^{III} =&-k\text{CS}_{N-N_s}[b]+\text{BF}[f;\text{Tr}_{N-N_s}\left(b\right)-N\tilde{A}_{1}-N\tilde{A}_{2}] -k \text{CS}_{N_s}[B] \\
&+Nk\text{BF}[\tilde{A}_{1};\tilde{A}_{2}]+Nk\text{CS}_{1}[\tilde{A}_{2}],\label{eq:phaseIII lsu}
\end{align}
After integrating out the Lagrange multiplier, we are left with
\begin{align}
 i\CL_{SU}^{III} =& -k\text{CS}_{N-N_{s}}[b^\prime]-k \text{CS}_{N_{s}}[B] \\
 &-\frac{Nk}{N-N_{s}}\left(N\text{CS}_{1}[\tilde{A}_{1}]+N_{s}\text{BF}[\tilde{A}_{1};\tilde{A}_{2}]+N_{s}\text{CS}_{1}[\tilde{A}_{2}]\right).
 \end{align}
The fact that we get such complicated Abelian Chern-Simons terms can be explained in a manner analogous to the non-Abelian $SU$ Higgsing discussed earlier. Indeed, as we should expect, this expression matches \eqref{eq:nab2 1}. More precisely, the complicated breaking of the $SU(N)$ field can be simplified by transforming into a $U(N)\times U(1)$ field and breaking down the $U(N)$ field, and the Lagrange multiplier encodes a change in coupling to \emph{both} Abelian factors $\tilde{A}_{1}$ and $\tilde{A}_{2}$. The splitting of the fiducial fermion modes once more occurs in a manner analogous to \eqref{eq:breaking_SU}.

\subsubsection*{$U$ Side}

Since $m_\Phi^2>0$ the $U(k)$ symmetry remains unbroken, but the dynamical fermions change the Chern-Simons terms. The change in Chern-Simons terms and edge modes follows in a manner practically identical to \eqref{eq:mdb f+f sector3B}.

\subsection*{Phase IVb}

This phase corresponds to $m_{\psi}>0$, $m_{\phi}^{2}<0$, $m_{\Psi}<0$, and $m_{\Phi}^{2}<0$. Additionally, this will be the first phase where we have to worry about singlet fermions, and we have $m_{s}>0$ in both theories.

\subsubsection*{$SU$ Side}

Similar to Phase III, the gauge group is spontaneously broken in this phase and this is slightly complicated by the fact this is the $SU$ side. Additionally, the dynamical and singlet fermions contribute additional Chern-Simons terms relative to Phase II, but they also contribute dynamical DWFs which makes said terms automatically non-anomalous.

\subsubsection*{$U$ Side}

Here the $U(k)$ symmetry is spontaneously broken to $U(k-N_f)\times SU(N_f)$, but the dynamical fermion behavior is the same as that of Phase II.  However, the singlet fermions have positive mass and thus shift a subset of the Chern-Simons level relative to that of Phase II. Although this is the first time we have seen the singlet fermion behaving differently from the dynamical fermions, there is nothing different about the way we end up at an non-anomalous theory. The singlet fermions give rise to DWFs which exactly compensate for their shift of the Chern-Simons levels in the bulk.

\subsection*{Phase IVa}

This phase corresponds to $m_{\psi}>0$, $m_{\phi}^{2}<0$, and $m_{s}<0$ on the $SU$ side and $m_{\Psi}<0$, $m_{\Phi}^{2}<0$, and $m_{s}<0$ on the $U$ side. Again, this phase is a repeat of what we have already looked at in Phase IVb but with negative mass singlet fermions. For the $U$ side, the singlet fermions have the same sign mass as the dynamical dynamical fermions and hence both contribute a shift to the Chern-Simons terms, but the \emph{different} masses break the flavor symmetry between the two.

\vspace{5mm}

Lastly, let us comment on the scalar boundary conditions for the master duality. As with the single species non-Abelian cases considered above, we can deduce whether $\phi$ and $\Phi$ obey Neumann or Dirichlet boundary conditions by comparing the global symmetry currents. Recall that when $N_s= 0$ the $\tilde{A}_2$ coupling vanished and the $\tilde{A}_1$ global symmetry could be attributed to the $U(1)_{m,b}$ symmetry. Meanwhile, when we took $N_f=0$ in Sec. \ref{sec:nab 2}, $\tilde{A}_1$ and $\tilde{A}_2$ could be combined into a new background field $\tilde{B}$ which was then associated with its own $U(1)_{m,b}$ symmetry. For the case when both $N_f$ and $N_s\ne 0$, the $\tilde{A}_1$ and $\tilde{A}_2$ background fields play the same roles. The combinations $\tilde{A}_1$ and $\tilde{A}_1+\tilde{A}_2$ are associated with two $U(1)_{m,b}$ symmetries, one whose $U(1)_b$ part is the $\psi$ matter, and the other, the $\phi$ matter. As such, all arguments of identifying global symmetries on either side of the duality to impose scalar boundary conditions still hold for the master duality, and so we find the same results, as shown in Table \ref{tab:n and d master}. 

\subsection{Generalization to $SO$ and $USp$}
\label{sec:so and usp}

Finally, we will briefly comment on the generalization of our methods to the versions of the master duality for the $SO$ and $USp$ groups in the presence of a boundary. In the bulk, these dualities are given by \cite{Benini:2017aed,Jensen:2017bjo}\footnote{Here we follow the notation of \cite{Aharony:2016jvv}, where $USp(2N)=Sp(N)$ and the levels of $SO$ groups are normalized to give Chern-Simons terms $\frac{k}{8\pi}\Tr\left(AdA-i\frac{2}{3}A^{3}\right)$. Also note that Majorana fermions come with an regularizing phase of $\exp(-i\pi\eta/4)$ instead of $\exp(-i\pi\eta/2)$, with $\eta$ the $\eta$-invariant.}
\begin{align}
SO(N)_{-k+\frac{N_{f}}{2}}\,\text{with \ensuremath{N_{s}} \ensuremath{\phi}\ and \ensuremath{N_{f}} \ensuremath{\psi}}\quad&\leftrightarrow\quad SO(k)_{N-\frac{N_{s}}{2}}\,\text{with \ensuremath{N_{f}} \ensuremath{\Phi}\ and \ensuremath{N_{s}} \ensuremath{\Psi}}\label{eq:master_so}\\
USp(2N)_{-k+\frac{N_{f}}{2}}\,\text{with \ensuremath{N_{s}} \ensuremath{\phi}\ and \ensuremath{N_{f}} \ensuremath{\psi}}\quad&\leftrightarrow\quad USp(2k)_{N-\frac{N_{s}}{2}}\,\text{with \ensuremath{N_{f}} \ensuremath{\Phi}\ and \ensuremath{N_{s}} \ensuremath{\Psi}}.\label{eq:master_usp}
\end{align}
Here, the matter is still in the fundamental representation of the respective gauge groups.  The difference now is that the scalars are real, and the fermions are Majorana. There are five massive phases following the same pattern as those considered for the $U/SU$ master duality. Note that the mass deformed phases match under the level-rank dualities generalized to the $SO$ and $USp$ cases \cite{Aharony:2016jvv},
\begin{align}
SO\left(N\right)_{-k}	&\leftrightarrow SO\left(k\right)_{N}\times SO\left(kN\right)_{-1}\\
USp\left(2N\right)_{-k}	&\leftrightarrow USp\left(2k\right)_{N}\times SO\left(4kN\right)_{-1}.
\end{align}

Accounting for the change to real fermions and scalars, there are half as many matter degrees of freedom as compared to the $U/SU$ dualities, which can most easily be understood by starting with complex scalars and Dirac fermions and imposing a reality condition \cite{Aharony:2016jvv}. Explicitly for the $USp$ duality, we will take $\psi$ to be a Dirac fermion but require that $\psi_{\alpha I}\Omega^{\alpha\beta}\tilde{\Omega}^{IJ}=(\psi^{\beta J})^c$; with $\psi^c$ the charge conjugate of $\psi$ and $\Omega^{\alpha\beta}$ ($\tilde{\Omega}^{IJ}$) symplectic invariant tensor of $USp(2N)$ ($USp(2N_f)$). Hence, integrating out real fermions provides half the change in Chern-Simons level as that of a full Dirac fermion.

As with the $U/SU$ case, the Chern-Simons terms are anomalous in the presence of a boundary. Fortunately, the fiducial fermion prescription used above can be generalized to be used with Majorana fermions. Alternatively, the fiducial Dirac fermions can still be used with the reality conditions discussed above. Thus, the $SO$ and $USp$ dualities can be rendered non-anomalous by rewriting Chern-Simons terms as fiducial Majorana fermions. Deriving the boundary conditions and DWFs for Majorana fermions follows similarly.

The global symmetries on either side of the master dualities also change slightly. For instance, the flavor symmetries of the fermions of the $SO$ ($USp$) duality are now $SO(N_f)$ ($USp(2N_f)$) on the left-hand side of \eqref{eq:master_so} and \eqref{eq:master_usp}, respectively. The fiducial Majorana fermions for a given $SO$ or $USp$ Chern-Simons term have an analogous ``flavor'' symmetry whose rank scales with the Chern-Simons level. Thus, when one chooses Dirichlet boundary conditions for the dynamical gauge field on one end of the duality, the fiducial fermions on the Neumann end once again share the same global symmetry on the boundary.

\section{Conclusion}\label{sec:conclusion}
Physical samples that we can drive to criticality and probe in a laboratory setting have boundaries, and too often conjectured dualities do not or cannot make explicit the role of boundary conditions. In order to understand what -- if any -- role dualities such as 2+1 dimensional master bosonization duality or any of its single species non-Abelian and Abelian limit cases play in describing physical critical systems, we must carefully analyze the admissible boundary theories consistent with bulk duality. Our previous work in building duality consistent boundary conditions where a prescriptive method for discovering the necessary edge modes was proposed was focussed solely on Abelian theories \cite{Aitken:2017nfd}.

The relative simplicity of the gauge sector in the Abelian dualities hid an important aspect of the choice of boundary conditions for the dynamical gauge fields.  In this work, we have reconciled the Abelian fiducial fermion prescription with those subtle aspects that are necessarily present in all non-Abelian bosonization dualities in 2+1 dimensions regardless of the types of fundamental matter considered.   The important takeaway is that the additional complication of having dynamical gauge fields on both sides of the duality necessitated an alternating prescription of boundary condition such that Neumann conditions are mapped to Dirichlet conditions across the duality.  As first observed in \cite{gaiotto} and later elaborated in \cite{Dimofte:2017tpi}, the reason this change in boundary conditions is due to emergent global symmetries in the boundary theories that must match in order to be duality-compatible.

Beyond simply analyzing the gauge sectors, in the preceding sections, we have constructed the necessary duality-compatible boundary conditions and additional edge modes for the master bosonization duality for Chern-Simons-matter theories  in \cite{Jensen:2017bjo, Benini:2017aed}.  A non-trivial check on the analysis in this work has been the consistent reduction of the duality-consistent boundary conditions in the master bosonization duality to the Abelian case.  The check furnished by the Abelian reduction also resolved a subtlety not addressed in \cite{Jensen:2017bjo, Benini:2017aed} regarding whether the Abelian gauge fields $U(1)_{m,b}$ and $U(1)_{F,S}$ were ordinary $U(1)$ or spin$_c$ connections.  Further, the motivation of the boundary conditions on the scalar sector of the $U$ side of the non-Abelian single species and master bosonization dualities discussed in \ref{sec:single-species-boundaries} provides a more satisfying picture than the Abelian analysis in \cite{Aitken:2017nfd} had suggested.  Lastly, the novel extension of the fiducial fermion prescription to $SO$ and $USp$ dualities filled out the spectrum of 2+1 dimensional bosonization dualities in the presence of a boundary.

That being said, there are further questions to ask in the context of 2+1 dimensional dualities involving Chern-Simons-matter theories in the presence of a boundary.  As noted at the start of this work, at the core of all of the bosonization dualities sits the basic level-rank duality familiar from WZW theories.  In the non-Abelian dualities, we \emph{cannot} integrate out the non-Abelian Chern-Simons terms for dynamical fields in the massive phases. Since the dynamical fields are related by the level-rank duality rather than simply being the same, this has resulted in slightly different boundary theories. One could then wonder whether WZW-matter theories participate in other non-trivial level-rank dualities.
To our knowledge, there has been little work done on the effects of level-rank duality for WZW theories with non-trivial matter sectors.

\subsubsection*{Acknowledgements}
We would like to thank Kristan Jensen for many helpful conversations during the course of this work.  The work of KA and AK was supported, in part, by the U.S.~Department of Energy under Grant No.~DE-SC0011637. The work of BR was funded, in part, by STFC consolidated grant ST/L000296/1.

\appendix
\section{Time-Reversed Master Duality}
In this appendix, we consider the time-reversed version of the master duality in order to explicate the subtlety of our conventions for the fermion mass terms.  That is, time-reversal acts as to change signs in the following way,
\begin{equation}
i\bar{\psi}\Dslash_{a}\psi\qquad\leftrightarrow\qquad i\bar{\psi}\Dslash_{a}\psi-i\left(\text{CS}_{1}[a]+2\text{CS}_\text{grav}\right).
\end{equation}
For example, on the $SU$ side of the master duality the fermion kinetic term becomes
\begin{align}
i\bar{\psi}\Dslash_{b+C-\tilde{A}_{1}}\psi\qquad\leftrightarrow\qquad & i\bar{\psi}\Dslash_{b+C-\tilde{A}_{2}}\psi-i\left(N_{f}\text{CS}_{N}[b^\prime]+N\text{CS}_{N_{f}}[C]\right)\nonumber \\
 & -i\left(NN_{f}\text{CS}_{1}[\tilde{A}_{2}]+2NN_{f}\text{CS}_\text{grav}\right)
\end{align}
This means the time-reversed master duality is given by
\begin{align}
\CL_{SU} & =|D_{b+B}\phi|^{2}+i\bar{\psi}\Dslash_{b+C-\tilde{A}_{2}}\psi-i\left(k\text{CS}_{N}[b]+\text{BF}\left[f;\text{Tr}_{N}\left(b-\mathds{1}_N(\tilde{A}_{1}-\tilde{A}_{2})\right)\right]\right)\nonumber \\
 & -i\left(-kN\text{BF}[\tilde{A}_{1};\tilde{A}_{2}]-kN\text{CS}_{1}[\tilde{A}_{2}]\right)+\CL_{\text{int}}\\
\CL_{U} & =\left|D_{c+C}\Phi\right|^{2}+i\bar{\Psi}\Dslash_{c+B-\tilde{A}_{2}}\Psi-i\left((N_s-N)\text{CS}_{k}[c]-N\text{BF}[\text{Tr}_{k}(c);\tilde{A}_{1}]\right)\nonumber \\
 & -i\left(k\text{CS}_{N_s}[B]+kN_{s}\text{CS}_{1}[\tilde{A}_2]+2k(N_s-N)\text{CS}_\text{grav}\right)+\CL_{\text{int}}^{\prime}.
\end{align}
where now the mass identification is
\begin{equation}
m_{\psi}\leftrightarrow m_{\Phi}^{2},\qquad m_{\phi}^{2}\leftrightarrow-m_{\Psi}.
\end{equation}
Note that signs of $\bar{\Psi}\Psi|\Phi|^2$ and $\bar{\psi}\psi|\phi|^2$ flip as well. The associated boundary conditions and edge modes are given in Table \ref{tab:n and d tr master}.
\begin{table}
\begin{centering}
\begin{tabular}{|>{\centering}p{5cm}|>{\centering}p{5cm}|>{\centering}p{5cm}|}
\hline
 & \textbf{$SU(N)$ Side} & \textbf{$U(k)$ Side}\tabularnewline
\hline

\textbf{Boundary Conditions} & $\psi^{+}_{\alpha I}|_{\partial}=0$

$\phi_{\alpha M}$: Neumann

$b^\prime$: Neumann & $\Psi^{-}_{\rho M}|_{\partial}=0$

$\Phi_{\rho I}$: Neumann

$c$: Dirichlet \tabularnewline

\hline

\textbf{Additional Edge Modes} &  $k\times\CL_{ff}^{+}[b^\prime+\tilde{A}_{1}\mathds{1}_{N}]$& {\bf{---}{---}{---}}\tabularnewline
\cline{2-3}
 & $Nk\times\CL_{ff}^{-}[0]$  & $Nk\times\CL_{ff}^{-}[0]$\tabularnewline

\hline
\hline
 & \textbf{$SU(N)$ Side} & \textbf{$U(k)$ Side}\tabularnewline
\hline

\textbf{Boundary Conditions} & $\psi^{+}_{\alpha I}|_{\partial}=0$

$\phi_{\alpha M}$: Dirichlet

$b^\prime$: Dirichlet & $\Psi^{-}_{\rho M}|_{\partial}=0$

$\Phi_{\rho I}$: Dirichlet

$c$: Neumann\tabularnewline

\hline

\textbf{Additional Edge Modes} &{\bf{---}{---}{---}} & $N\times\CL_{ff}^{-}[c^{\prime}+\tilde{c}+\tilde{A}_{1}\mathds{1}_{k}]$  \tabularnewline
\cline{2-3}
 &  $Nk\times\CL_{ff}^{+}[\tilde{A}_{1}]$ & $Nk\times\CL_{ff}^{+}[\tilde{A}_{1}]$\tabularnewline
\cline{2-3}
 & $Nk\times\CL_{ff}^{-}[0]$  & $Nk\times\CL_{ff}^{-}[0]$\tabularnewline

\hline
\end{tabular}
\par\end{centering}
\caption{The top (bottom) table counts the additional edge modes when choosing Neumann and Dirichlet boundary conditions on the dynamical gauge fields in $SU$ ($U$) and $U$ ($SU$) side respectively when $N_f\neq0$ and $N_s\neq 0$ for the time-reversed master duality. \label{tab:n and d tr master}}
\end{table}

\bibliographystyle{JHEP}
\bibliography{master_dual_boundaries_bib}
\end{document}